\documentclass[twocolumn]{aastex62}

\newcommand{\ts}{\textsuperscript}

\usepackage{xcolor}
\usepackage{float}
\usepackage[nolist]{acronym}

\usepackage[pagewise]{lineno}

\graphicspath{{./}{figures/}}

\accepted{May 5, 2019}

\shorttitle{Unsupervised Clustering of Convolutionally Encoded Radio-astronomical Images}
\shortauthors{Ralph et al.}

\begin{document}

\begin{acronym}[MMMMM]

\acro{Ae}{Autoencoder}
\acro{Adam}{Adaptive Moment}
\acro{AGN}{Active Galactic Nuclei}
\acro{ASKAP}{Australian Square Kilometre Array Pathfinder}
\acro{ATCA}{Australia Telescope Compact Array}

\acro{BMU}{Best Matching Unit}
\acro{CAe}{Convolutional Autoencoder}
\acro{CNN}{Convolutional Neural Network}
\acro{CPU}{Central Processing Unit}
\acro{CSV}{Comma-Separated Value}
\acro{CTA}{Cherenkov Telescope Array}
\acro{DBM}{Deep Belief Machine}
\acro{DR1}{Data Release 1}
\acro{DT}{Decision Tree}
\acro{ELT}{Extremely Large Telescope}
\acro{EMU}{Evolutionary Map of the Universe}
\acro{FIRST}{Faint Images of the Radio Sky at Twenty Centimetres}
\acro{FITS}{Flexible Image Transport System}
\acro{GAN}{Generative Adversarial Network}
\acro{GNG}{Growing Neural Gas}
\acro{GPU}{Graphics Processing Unit}
\acro{HC}{Hierarchical Clustering}
\acro{HMM}{Hidden Markov Model}
\acro{AUI}{Associated Universities, Inc}
\acro{ID}{Identification Number}
\acro{IR}{Infrared}
\acro{ISM}{Inter-Stellar Medium}
\acro{KNN}{k-Nearest Neighbour}
\acro{LOFAR}{Low-frequency Array}
\acro{LOTSS}{\ac{LOFAR} Two-metre Sky Survey}
\acro{LReLU}{Leaky Rectified Linear Unit}
\acro{LSST}{Large Synoptic Survey Telescope}
\acro{LSTM}{Long Short-Term Memory}
\acro{MLP}{Multi-Layer Perceptron}
\acro{MNIST}{Modified National Institute of Standards and Technology}
\acro{MSE}{Mean Square Error}
\acro{MRO}{Murchison Radio-astronomy Observatory}
\acro{ML}{Machine Learning}
\acro{MLP}{Multi-Layer Perceptron}
\acro{NaN}{Not a Number}
\acro{NN}{Neural Network}
\acro{NRAO}{National Radio Astronomy Observatory}
\acro{NVSS}{NRAO VLA Sky Survey}
\acro{PCA}{Principle Component Analysis}
\acro{PB}{Petabyte}
\acro{PINK}{Parallelized rotation and flipping Invariant Kohonen maps}
\acro{PNG}{Portable Network Graphic}
\acro{PSF}{Point Spread Function}
\acro{RAM}{Random Access Memory}
\acro{RBM}{Restricted Boltzmann Machine}
\acro{ReLU}{Rectified Linear Unit}
\acro{RGZ}{Radio Galaxy Zoo}
\acro{RMS}{Root Mean Squared}
\acro{RNN}{Recurrent Neural Network}
\acro{RTF}{Random Tree Forest}
\acro{SD}{Standard Deviation}
\acro{SDSS}{Sloan Digital Sky Survey}
\acro{SED}{Spectral Energy Distribution}
\acro{SETI}{Search of Extraterrestrial Intelligence}
\acro{SKA}{Square Kilometre Array}
\acro{SNR}{Supernova Remnants}
\acro{SOM}{Self-Organising Map}
\acro{SVM}{Support Vector Machine}
\acro{TB}{Terabyte}
\acro{t-SNE}{t-Distributed Stochastic Neighbour Embedding}
\acro{UMAT}{unified distance matrix}
\acro{VAe}{Variational Autoencoder}
\acro{VLA}{Very Large Array}
\acro{VLBI}{Very Long Base Line Interferometry}
\acro{WISE}{Wide-field Infrared Survey Explorer}
\acro{WTF}{Widefield Outlier Finder}
\end{acronym}

\title{Radio Galaxy Zoo: Unsupervised Clustering of Convolutionally Auto-encoded Radio-astronomical Images}

\author{Nicholas O. Ralph}
\affiliation{Western Sydney University, \\
School of Computing, Engineering and Mathematics\\
Locked Bag 1797, Penrith NSW 2751, Australia}
\affiliation{CSIRO
Astronomy and Space Science,
\\Australia Telescope National Facility,
\\PO Box 76, Epping, NSW 1710, Australia}

\author{Ray P. Norris}
\affiliation{Western Sydney University, \\
School of Computing, Engineering and Mathematics\\
Locked Bag 1797, Penrith NSW 2751, Australia}
\affiliation{CSIRO
Astronomy and Space Science,
\\Australia Telescope National Facility,
\\PO Box 76, Epping, NSW 1710, Australia}

\author{Gu Fang}
\affiliation{Western Sydney University, \\
School of Computing, Engineering and Mathematics\\
Locked Bag 1797, Penrith NSW 2751, Australia}

\author{Laurence A. F. Park}
\affiliation{Western Sydney University, \\
School of Computing, Engineering and Mathematics\\
Locked Bag 1797, Penrith NSW 2751, Australia}

\author{Timothy J. Galvin}
\affiliation{CSIRO Astronomy and Space Science \\
Kensington WA 6151, Australia}

\author{Matthew J. Alger}
\affiliation{Research School of Astronomy and Astrophysics,\\ The Australian National University,\\ Canberra, ACT 2611, Australia}
\affiliation{Data61, CSIRO, Canberra, ACT 2601, Australia}

\author{Heinz Andernach}
\affiliation{Departamento de Astronom\'ia, DCNE,\\ Universidad de Guanajuato, Apdo. \\Postal 144, CP 36000, Guanajuato, Gto., Mexico}

\author{Chris Lintott}
\affiliation{Department of Physics, University of Oxford,\\Denys Wilkinson Building, Keble Road, Oxford,  OX1 3RH, UK}

\author{Lawrence Rudnick}
\affiliation{Minnesota Institute for Astrophysics, University of Minnesota}

\author{Stanislav Shabala}
\affiliation{University of Tasmania, School of Natural Sciences,\\ Private Bag 37, Hobart, Tasmania 7001, Australia}

\author{O. Ivy Wong}
\affiliation{International Centre for Radio Astronomy (ICRAR),\\ M468, The University of Western Australia,\\ 35 Stirling Highway, Crawley, WA 6009, Australia}


\begin{abstract}
This paper demonstrates a novel and efficient unsupervised clustering method with the combination of a \ac{SOM} and a convolutional autoencoder. The rapidly increasing volume of radio-astronomical data has increased demand for machine learning methods as solutions to classification and outlier detection. Major astronomical discoveries are unplanned and found in the unexpected, making unsupervised machine learning highly desirable by operating without assumptions and labelled training data. Our approach shows \ac{SOM} training time is drastically reduced and high-level features can be clustered by training on auto-encoded feature vectors instead of raw images. Our results demonstrate this method is capable of accurately separating outliers on a \ac{SOM} with neighborhood similarity and K-means clustering of radio-astronomical features complexity. We present this method as a powerful new approach to data exploration by providing a detailed understanding of the morphology and relationships of \ac{RGZ} dataset image features which can be applied to new radio survey data.

\end{abstract}


\newpage

\section{Introduction} \label{sec:intro}

Large radio continuum surveys have played a key role in our understanding of the evolution of galaxies \citep{norris2017extragalactic}.
Exceptionally large surveys such as \ac{LOTSS} \citep{2017A&A...598A.104S} and the Evolutionary Map of the Universe \citep[EMU]{norris2011emu} are expected to detect 30 million and 70 million radio sources respectively. The sheer scale and complexity of these datasets is pushing researchers towards automated techniques such as machine learning with \acp{NN}.   

\acp{NN} are networks of functions termed "neurons" that operate as function approximators. A typical implementation of \acp{NN} include the \ac{MLP}, as a class of feed-forward \acp{NN} with multiple neuron layers. In these \acp{MLP}, neuron parameters are typically learned via backpropagation \citep{1100330}, where weights are updated using gradient descent of a given loss function as a difference measure between target and predicted output. \\ 
A \ac{NN} trained to classify images with a specific orientation and scale will, however, encounter difficulties when classifying the same training image at an untrained angle or scale \citep{perantonis1992translation}. Affine transformations such as rotation, scaling and translation, are a common cause of machine learning prediction errors. A classical solution involves augmenting a training set with random rotations and scaling at the cost of training time. Alternatively, a network can be made invariant to scaling by adding convolutional and max-pooling layers. Rotational invariance is more easily solved with the addition of rotated training images.

\acp{NN} such as SkyNet \citep{doi:10.1093/mnras/stu642} have accurately classified astronomical data using supervised learning of pre-classified examples. Efforts to use these supervised neural networks have been supported with citizen science projects such as the \acl{RGZ} \citep{banfield2015radio} which has created large labelled datasets of radio sources. This \ac{RGZ} dataset has been used to successfully train classifiers for source classification \citep{wu2018radio,lukic2018radio} and radio source host galaxy cross-identification \citep{alger2018radio}. However, this supervised training is not always suitable in outlier detection and separating source complexity as it requires a more complete knowledge of all potential classes of new unseen data. Given that most of the major discoveries in astronomy have been unplanned \citep{norris2017discovering}, this is a major shortcoming. 

\begin{figure}[!ht]
        \includegraphics[width=0.47\textwidth]{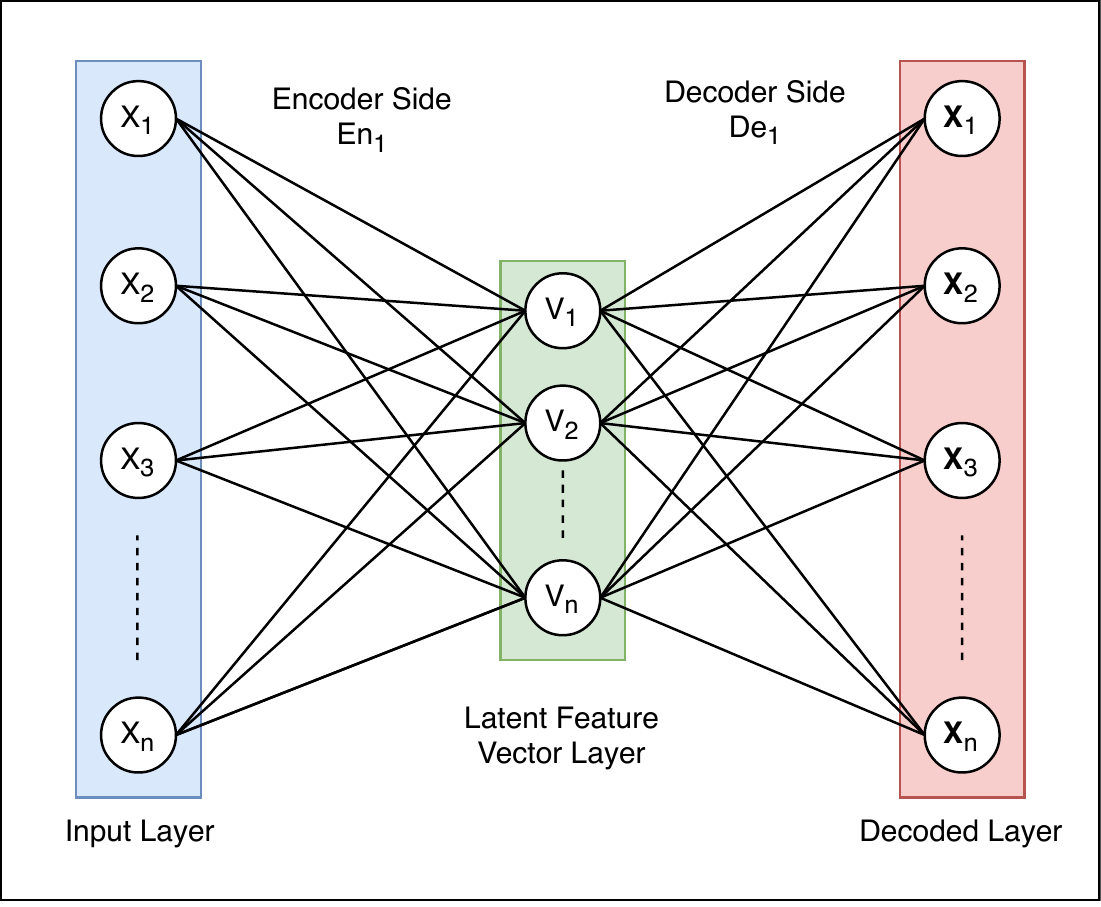}
    \caption{The network configuration of a simple fully-connected \ac{NN} variant autoencoder, featuring the encoder input layer, the down-sampled latent feature vector layer and the reconstructed decoder output layer.}\label{fig: basic-ae}
\end{figure}

Unsupervised learning techniques bridge this gap by working with no assumptions about input data. An autoencoder is an example of unsupervised learning, designed for dimensionality reduction. Autoencoders work by extracting and compressing the features of input images into a feature vector \citep{sanger1989optimal}. The ideal autoencoder is trained to perfectly compress and restore input data with no loss. The layer configuration of a typical \ac{NN} autoencoder variant (as shown in Figure \ref{fig: basic-ae}) uses an \ac{MLP} architecture with backpropagation learning to reduce input data to a compact feature representation on the encoding side before returning it to its original form on the decoding side. The layer configuration of the encoder and decoder are usually very similar. Autoencoder prediction loss is given as the difference between input data and the decoded output. This loss is naturally an indicator of the performance of the network but is also sensitive to differences between an input image and the training set. Since loss is calculated from the input data, a label set is not required and the network can be trained unsupervised. Autoencoders have seen success in many image processing applications with the addition of convolution, max pooling and denoising architecture \citep{xie2012image}. 

Abstract relationships and topology in large datasets can be interpreted by visualising auto-encoded feature vectors with dimensionality reduction methods such as \acl{PCA} \citep[PCA]{hotelling1933analysis} onto a learning manifold. More complex approaches such as \acl{SOM} \citep[SOM]{kohonen1997exploration} have been recognised as especially powerful unsupervised data exploration tools in astronomy \citep{polsterer2015automatic, tasdemir2009exploiting}. By adapting to the shape of encoded latent vectors, these \acp{SOM} can display various topological relationships and morphology distributions. Moreover, these algorithms have been augmented to produce labelled classification and source separation with K-means clustering \citep{lloyd1982least}.

This paper demonstrates a novel and efficient unsupervised clustering by combining a self-organising map with a convolutional autoencoder as a variation of the \ac{CNN}. Using our proposed method, we show that \ac{SOM} training time is drastically reduced by training on the compressed autoencoder feature vectors of the \ac{RGZ} image. This method is demonstrated as a powerful data exploration and visualisation tool. This approach shows K-means clustering of trained \ac{SOM} weights as a method of grouping radio-astronomical features in these \ac{RGZ} images and effectively separating sources by their complexity. We demonstrate our method as an effective and efficient solution to understanding the morphology and relationships of \ac{RGZ} images that can be applied to unexplored fields for discovery purposes. This approach is in contrast to typical \ac{CNN} applications in astronomy such as \citet{gravet2015catalog}, which instead create a system for morphological prediction and classification without this element of exploration. The use of abstracted image representations as auto-encoded feature vectors are a significant novel aspect of our method and offer great advantages in computational efficiency compared to this prior work and other \ac{CNN} implementations such as the system in \citet{dieleman2015rotation}, which also uses random rotation training augmentations for rotational invariance, but are trained on complete images alone.

\section{Data} \label{sec:style}
\acl{RGZ} is a citizen science project for radio image classification by volunteers via web interface \citep{banfield2015radio}. The majority of the radio image data in Radio Galaxy Zoo comes from the 1.4 GHz \ac{FIRST} survey catalogue \citep{becker1995first} version 14 March 2004. \ac{FIRST} covers over 9000 square degrees of the northern sky down to a 1~$\sigma$ noise level of 150~$\mu$Jy~beam$^{-1}$ at 5$''$ resolution. We use a total of 80,000 \ac{FIRST} images from the \acs{RGZ} Data Release 1 catalog \citep{wong} in this paper. 

Hand labelled \ac{RGZ} annotations of the dataset contain the number of components for every resolved source in the image \citep{banfield2015radio}. These annotations also include the number of brightness peaks above a set threshold within an image. We have encoded these labels as components-peaks, e.g single component with a single peak is 11, two components with two peaks is 22. Table \ref{table: dataset_population} shows that the largest fraction of the dataset contains single component single peak sources. 

  \begin{table}[h!]
    \begin{center}
      \begin{tabular}{lcc}
        \hline
RGZ Label &  Population Division &   Category \\
          \hline
       11 &               0.6110 &     Simple \\
       12 &               0.1533 &    Complex \\
       13 &               0.0153 &    Complex \\
       14 &               0.0020 &  Anomalous \\
       15 &               0.0003 &  Anomalous \\
       16 &               0.0001 &  Anomalous \\
       22 &               0.1438 &    Complex \\
       23 &               0.0195 &    Complex \\
       24 &               0.0028 &  Anomalous \\
       33 &               0.0340 &    Complex \\
       34 &               0.0053 &  Anomalous \\
       35 &               0.0008 &  Anomalous \\
       36 &               0.0003 &  Anomalous \\
       44 &               0.0068 &  Anomalous \\
       45 &               0.0014 &  Anomalous \\
       46 &               0.0004 &  Anomalous \\
       55 &               0.0020 &  Anomalous \\
       56 &               0.0005 &  Anomalous \\
       57 &               0.0002 &  Anomalous \\
       67 &               0.0002 &  Anomalous \\
        \hline
      \end{tabular}
            \caption{\acs{RGZ} \acs{DR1} classes by population. All point sources (RGZ label 11) are categorised as simple, all sources labelled as having more than one component or peak as complex, and any source with more than 3 components or peaks as anomalous.  
                \label{table: dataset_population}}
    \end{center}
  \end{table}

\section{Image Preprocessing}

\begin{figure*}
  \gridline{\fig{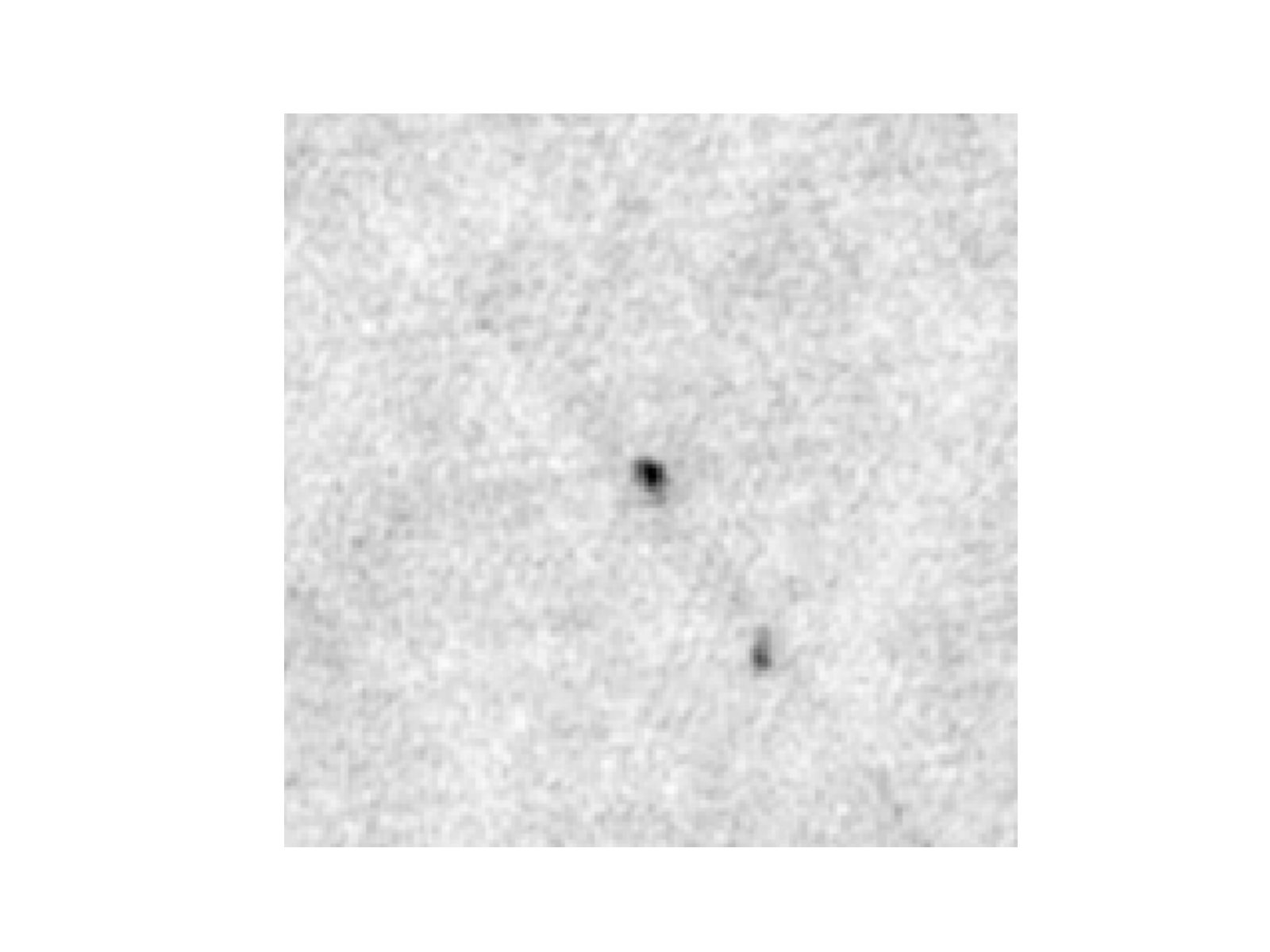}{0.32\textwidth}{(a)}
            \fig{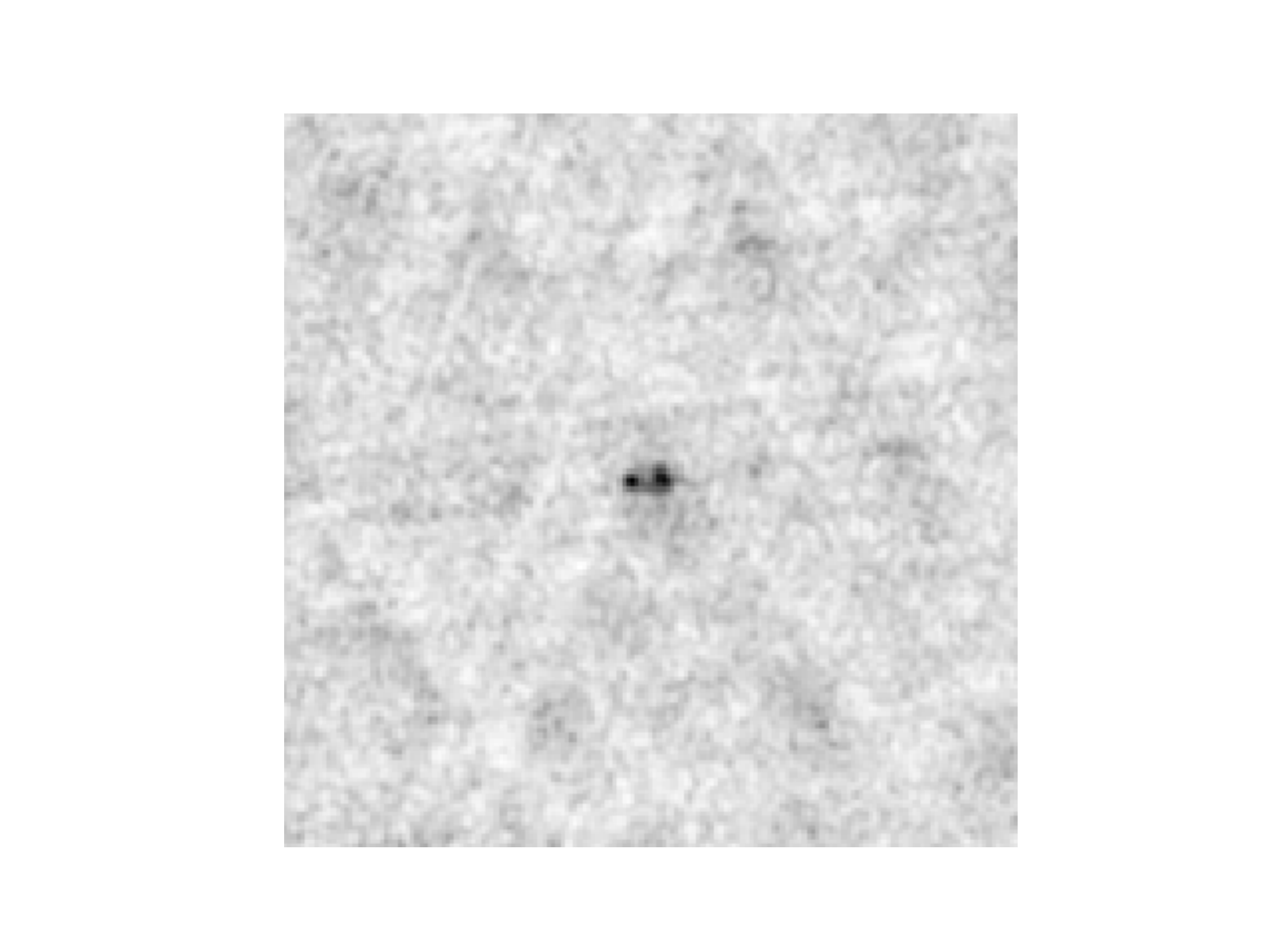}{0.32\textwidth}{(b)}
            \fig{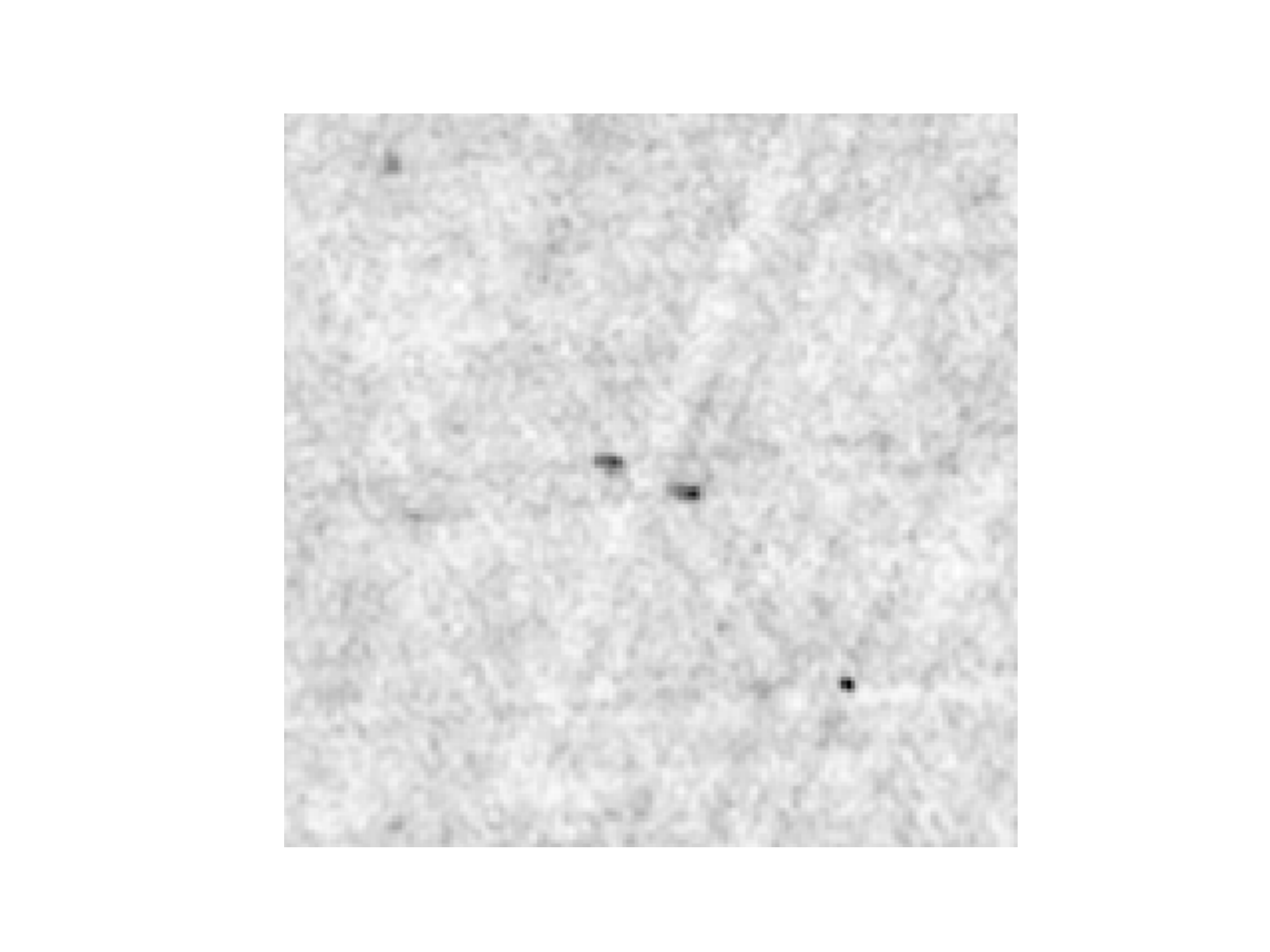}{0.32\textwidth}{(c)}
            }
  \gridline{\fig{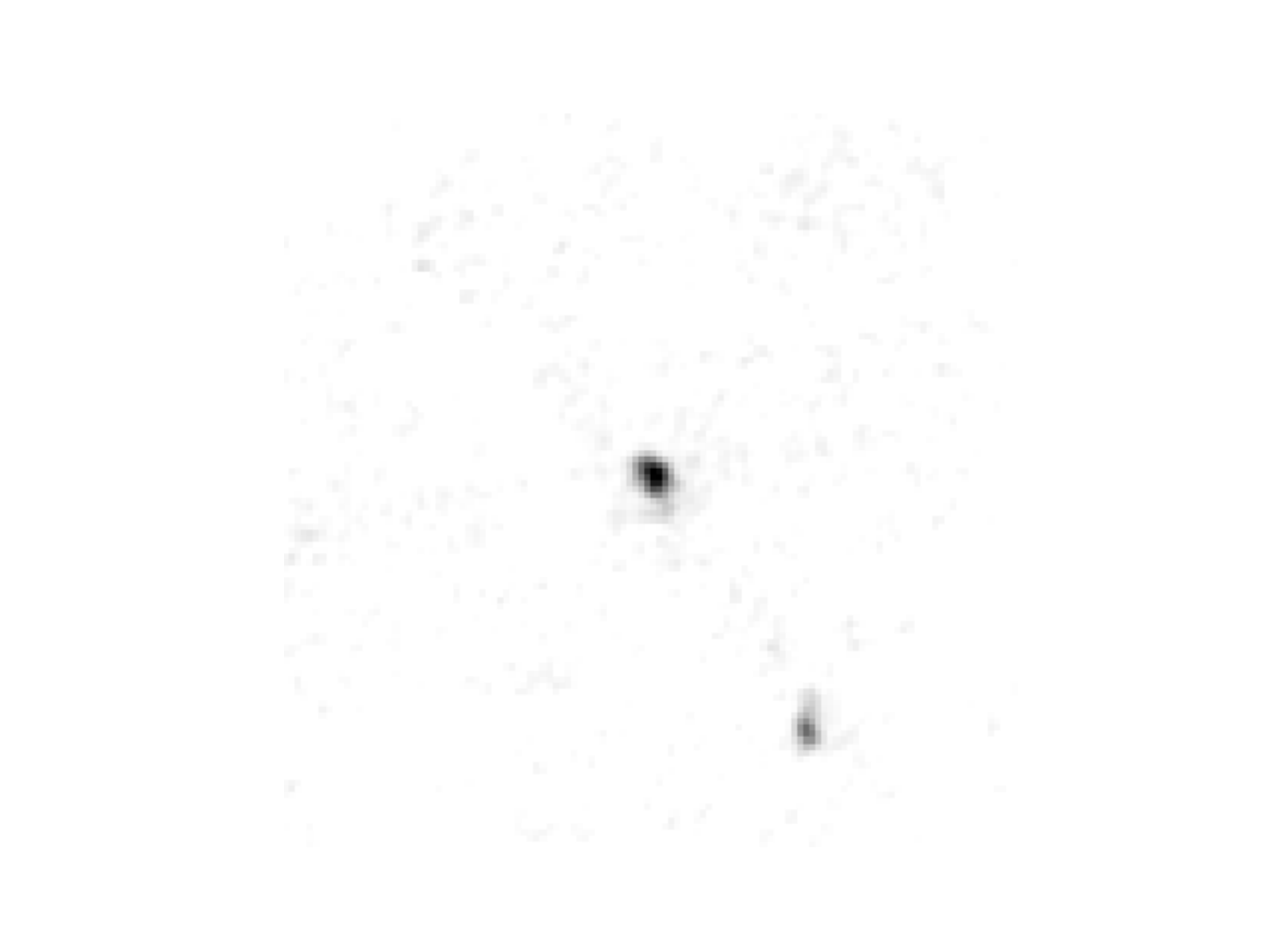}{0.32\textwidth}{(d)}
            \fig{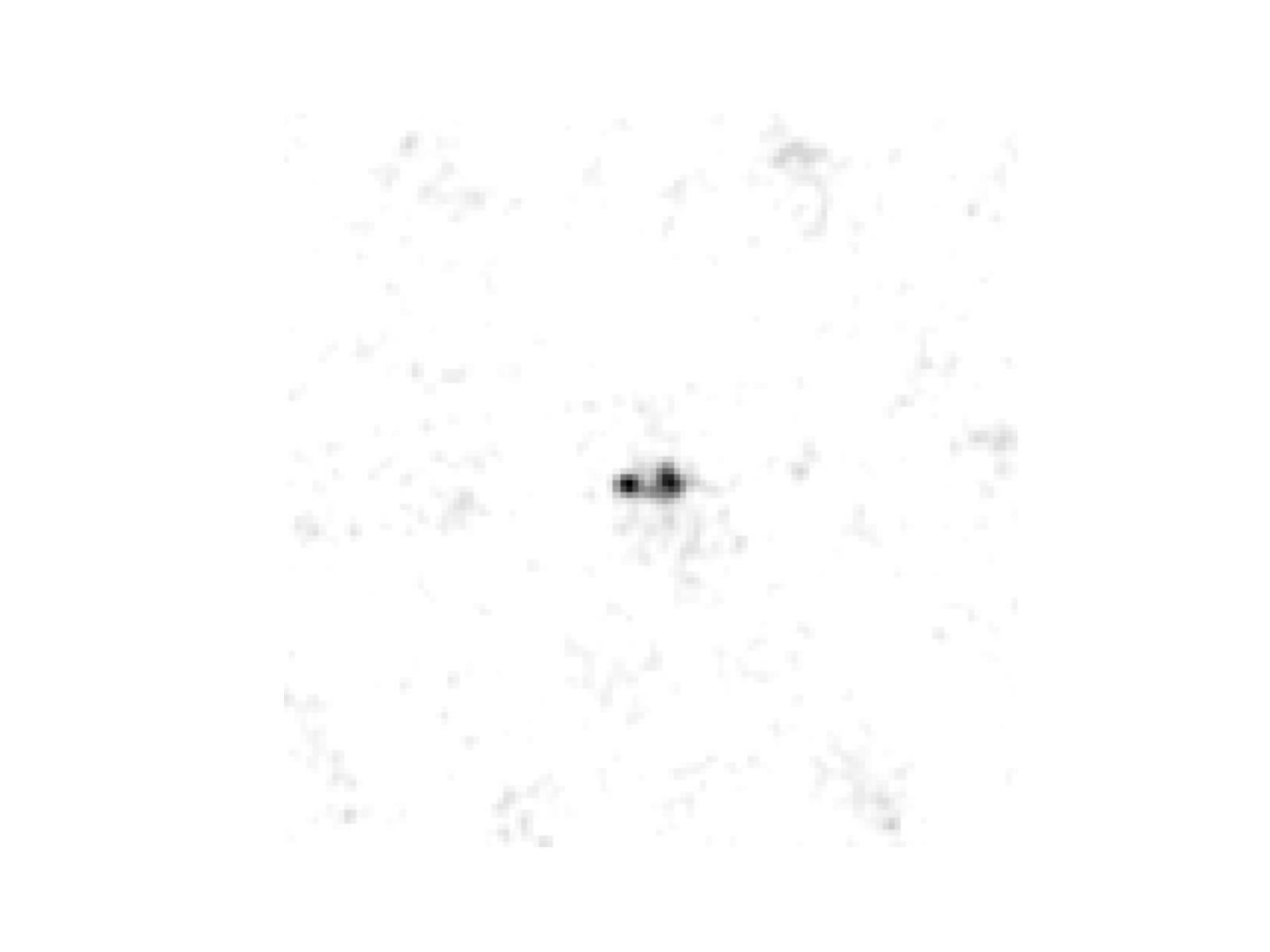}{0.32\textwidth}{(e)}
            \fig{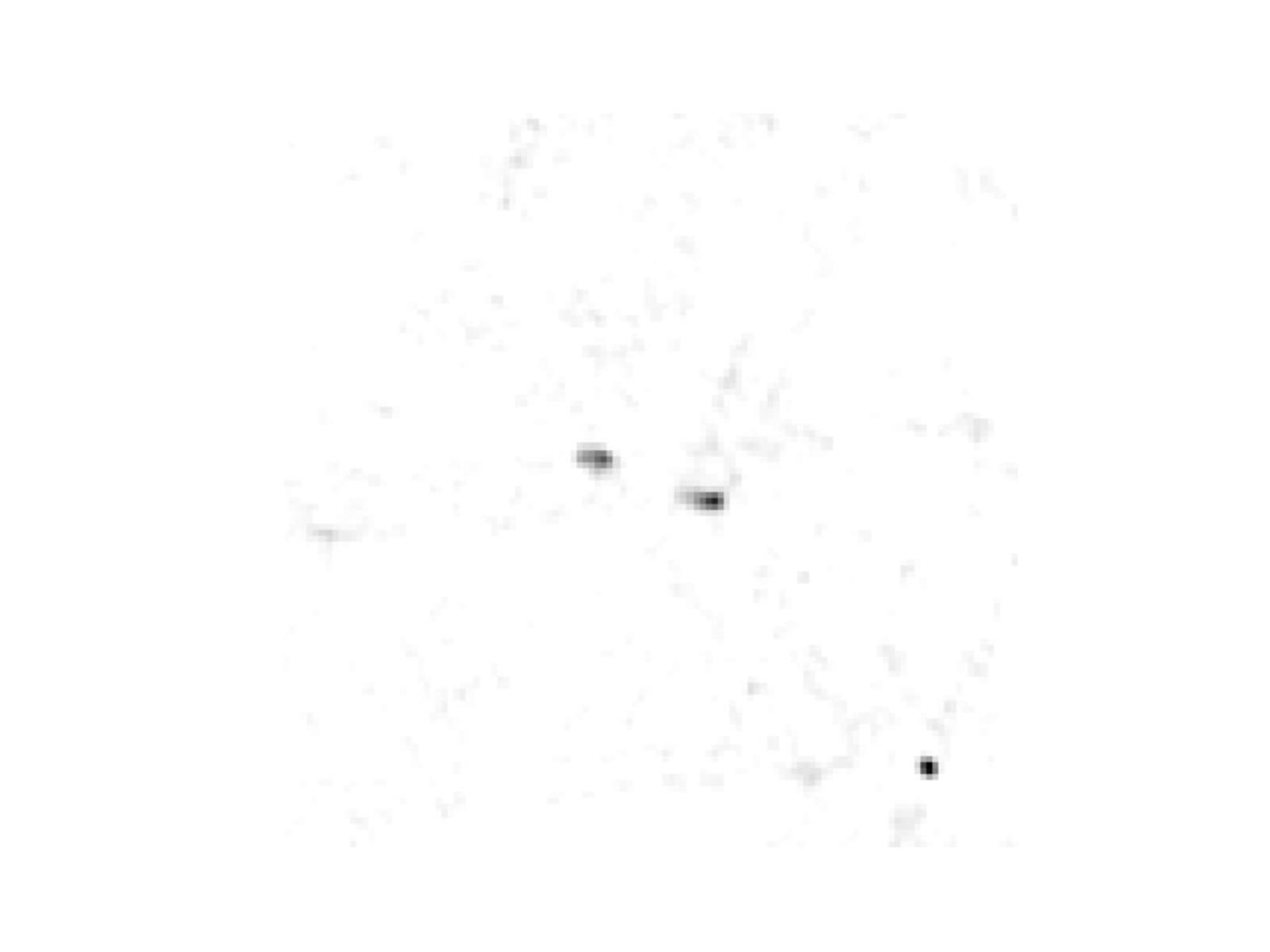}{0.32\textwidth}{(f)}
            }
            
    \caption{RGZ image preprocessing, (a),(b) and (c) as unprocessed \ac{FIRST} images. Images (d), (e) and (f) shown as the preprocessing output with noticeable improvement to noise, where a large portion of the background pixels are set to zero}. \label{fig:f2}
\end{figure*}

Radio images are contaminated by remnants of the instrument's \ac{PSF}. This contamination is often a major component of the feature space of the \ac{RGZ} training set. These elements must be removed, as we found that without proper preprocessing, clustering resulted in two classes: "noisy" and "not noisy", distinguished only by intensity distribution. 
Early preprocessing methods used in this investigation were effective at removing noise but had a tendency to remove faint sources and produce artifacts. As a result, we adopted the preprocessing method of \cite{galvin:pink-inpress} with results shown in Figure \ref{fig:f2}. This approach corrects blank pixels in images at the edge of the \ac{FIRST} image mosaic, sigma clips and normalises pixel intensity using the following procedure: 

\begin{enumerate}

\item Blank pixel regions found in images close to the edge of the \ac{FIRST} mosaic are corrected. This correction replaces these masked values with a random sample of the mean and standard deviation of valid pixels around the outer edge region of the image (assuming a normal distribution). These samples are extracted from the outer regions of the image with few astronomical features to properly sample the background noise.

\item Noise is removed and background flux is corrected with sigma clipping. This operation subtracts the mean background pixel value and clips all pixel intensities below 1~$\sigma$ to zero. 

\item Intensity scaling is applied to normalise the global intensity of each image.

\item All images are additionally cropped for the purposes of this paper, from $132\times132$ pixels to the centre 120x120 pixels to reduce the dataset size while preserving salient features.  

\end{enumerate}

\section{Method}
In this section, we outline our method of reducing \ac{RGZ} images with convolutional autoencoding to a compact feature vector for clustering and visualisation using a \ac{SOM} and K-means clustering. These methods were developed using the Python Language using a 24 core Intel(R) Xeon(R) \ac{CPU} E5-2650 v4 at 2.20GHz. We implemented our system using a number of Python machine learning packages. The Google Tensorflow Machine Learning Library \citep{abadi2016tensorflow} was used to create the autoencoder network, and Somoclu library \citep{wittek2013somoclu} was used to implement the \ac{SOM}.

\subsection{Affine Invariant Convolutional Autoencoders}
  In our method, we extract the latent relationships of \ac{RGZ} image features using a convolutional autoencoder.  

  We use a convolutional autoencoder with three convolutional layers trained on a random sample of 10,000 images and validated on a separate set of 10,000 images. Table \ref{table: final_ae_config} outlines the implemented \ac{NN} autoencoder with a \ac{MLP} architecture. The architecture was chosen experimentally through a brief set of trials to determine the best performing configuration. All convolutional layers use the \acl{LReLU} activation function \citep[LReLU]{maas2013rectifier} given its demonstrated success \citep{lecun_deep_2015}. All activation functions in this network use an activation function slope of 0.2. Additionally, the \acl{Adam} \citep[Adam]{kingma_adam:_2014} optimiser was chosen with a Tensorflow default learning rate of 0.01 due to its considerable use and success as a simple, computationally efficient and effective method in training large networks \citep{ruder_overview_2016}. We use a small batch size of 16 images during training. Small batch sizes in this range have been previously shown to allow autoencoder training to converge on solutions faster than larger batch sizes \citep{8323035}
  
  The encoder output layer here is a max-pooling operation, with the decoder input layer restoring the latent vector to its dimensions before max pooling with a linear interpolator. A latent vector with a 900x1 shape is the consequence of the number and dimensions of kernels used in the network. These dimensions can be modified by scaling the input image, however, it was found that training converged quickly with this 900x1 latent vector. The dimensions of this vector represent a significant reduction to the original image dimensions (120x120, 14400x1) while still containing sufficient free parameters to preserve information for decompression with minimum error. 

\begin{table*}[h!t]
  \begin{center}
    \caption{Outline of final optimised autoencoder architecture and layer configuration, where all convolutional layers use \ac{LReLU} activation functions} \label{table: final_ae_config}

    \begin{tabular}{lccccc }
       \hline
 Network Section & Layer & Function &  Input  &Filter Size &  Stride \\
       \hline
         Encoder & 0 & Input  & $120\times120\times1$ & - & -\\
                  & 1 & Convolution 1  & $120\times120\times1$ & $5\times5\times1$ & $1\times2\times2\times1$ \\
                 & 2 & Convolution 2  & $60\times60\times1$ & $5\times5\times32$ & $1\times2\times2\times1$ \\
                 & 3 & Convolution 3  & $30\times30\times1$ & $5\times5\times1$ & $1\times2\times2\times1$ \\
                 & 4 & Max-Pool 1 & $30\times30\times1$ & $3\times3\times1$ & $1\times1\times1\times1$ \\
                        \hline
         Centre  & 5 & Latent Vector  & $30\times30\times1$ & - & - \\
                \hline
         Decoder & 6 & De-Pool 1  & $30\times30\times1$ & $3\times3\times1$ & $1\times1\times1\times1$ \\
                  & 7 & Convolution 4  & $30\times30\times1$ & $5\times5\times1$ & $1\times2\times2\times1$ \\
                 & 8 & Convolution 5  & $60\times60\times1$ & $5\times5\times32$ & $1\times2\times2\times1$ \\
                 & 9 & Convolution 6  & $120\times120\times1$ & $5\times5\times1$ & $1\times2\times2\times1$ \\
                   & 10 & Output  & $120\times120\times1$ & - & -\\
          \hline

     \end{tabular}
  \end{center}
\end{table*}

Loss is calculated as the pixel \ac{MSE} between the autoencoder prediction image and the input image, averaged across the batch. We investigate rotational invariance by also training on images randomly rotated during training. This rotational invariance ideally prevents clustering methods from recognising rotation as a feature distinguished enough to separate it from its class. As the autoencoder is still being trained on rotation, these features will still be encoded into the latent vectors but with less weight.    

\begin{figure*}[!ht]
        \includegraphics[width=\textwidth]{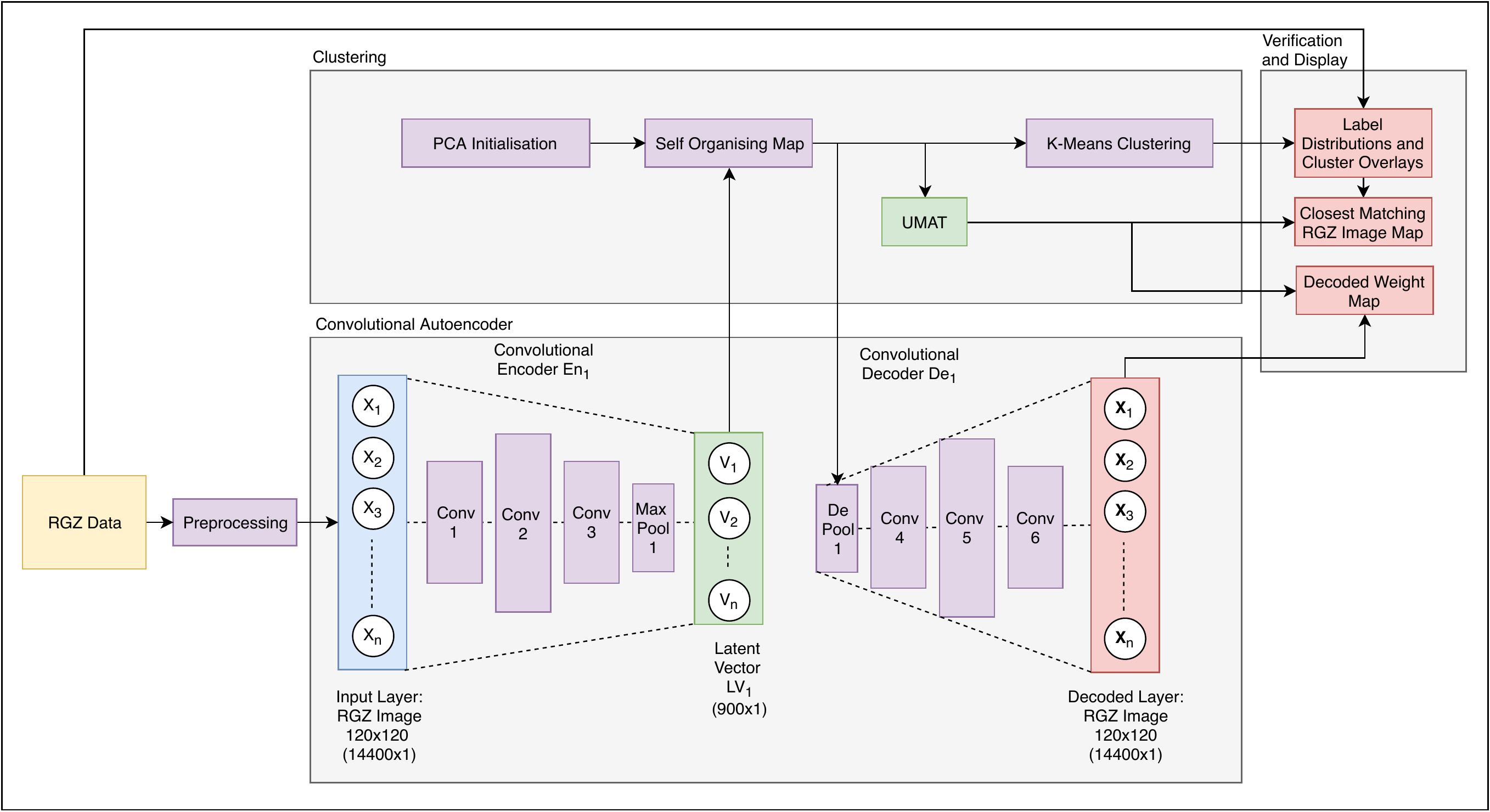}
    \caption{Overall pipeline configuration with the convolutional autoencoder architecture and self-organising map used in this paper. This network compresses input images with 14400 elements to the latent feature vector with 900 elements for clustering. Encoder and decoder architecture is identical with three convolutional layers at 1, 32 and 1 layers deep, selected experimentally. The \ac{SOM} weights are initialised using \ac{PCA} and trained on the encoded latent vectors. Learned \ac{SOM} weights are reconstructed using the decoder network of the autoencoder to display an approximation of the learned radio-astronomical features. K-means clustering is applied to \ac{SOM} weights and labeled for verification to compare the map clusters against \ac{RGZ} labels.} \label{fig:ae_architecture_full}
\end{figure*}

\subsection{Self Organising Maps}

  Self-Organising Maps (SOM) are data analysis methods used in unsupervised clustering and data exploration. SOMs create similarity maps or learning manifolds of input data where distinct groups of neurons reflect latent clusters in the data. A \ac{SOM} models datasets by iteratively updating a grid of neuron weight vectors $m_t$. This is achieved by moving toward similar data points $x(t)$ on the \ac{SOM} manifold by refining neurons weights with a neighbourhood distance function $h_{ci}$ of each neuron $i$, by a decaying learning rate $\alpha$ which is balanced to let all neurons stabilise in optimal time, as characterised in Equation \ref{eqn:method-som-weight}. A well-trained \ac{SOM} after $m$ epochs will visualise the distribution of the input \ac{RGZ} training data as various high-level topological relationships and morphology distributions.

\begin{equation} 
    \label{eqn:method-som-weight}  
m_{i}(t+1)=m_{i}(t) + \alpha(t) \cdot h_{ci}(t)[x(t) - m_i(t)] 
\end{equation} 

We trained our \ac{SOM} on a random sample of 30,000 autoencoder latent vectors and validated with a separate 30,000 latent vectors. Neither set includes encoded \ac{RGZ} images used in autoencoder training or validation to ensure the system remains relatively generalisable. This training was conducted with a focus on efficiency and demonstrating reasonable clustering using the following procedure:


\begin{enumerate}

  \item Initialise \ac{SOM} grid neurons with a \ac{PCA} learning manifold of the latent feature vector set. This approach allows the \ac{SOM} to model an already delineated \ac{PCA} space. 
  
  \item Select a random latent feature vector from the training set. 

  \item Locate \ac{BMU} neurons as the `closest' neuron to the selected data point. A common and reliable distance metric used to calculate this is Euclidean distance.


  \item Move all neuron weights within the neighbourhood toward the data point by updating neuron weights $m_i(t)$ as a function of neighbourhood function $h_{ci}(t)$, decay function $\sigma$, and learning rate $\alpha(t)$, as shown in Equation \ref{eqn:method-lin_som_learning_radius}. This neighbourhood function can be represented with several shapes by their radius, namely: 
  
  Linear:
  \begin{equation} 
    \label{eqn:method-lin_som_learning_radius}  
        h_{ci} = h_{c0}\sigma
  \end{equation} 
 
 and Gaussian:
    \begin{equation} 
        \label{eqn:method-exp_som_learning_radius}  
            h_{ci} = e^{-\frac{D(n_{c}, n_{i})}{2\sigma^2}}
      \end{equation}

  where exponential decay is given by:
  \begin{equation} 
    \label{eqn:method-exp_som_learning_radius}  
        \sigma_{exp} = e^{-\frac{t}{\tau}}
  \end{equation} 
  
  and linear decay:
  
  \begin{equation} 
    \label{eqn:method-exp_som_learning_radius}  
        \sigma_{linear} = -\frac{t}{\tau}
  \end{equation} 

 where $\tau$ is a decay constant, usually given as the number of training epochs. $D(x,y)$ is given as the distance function (Euclidean distance for the purposes of this paper) between the weight vector of the current excited neuron $n_{c}$ and weight vector of the winning neuron $n_{i}$ on the \ac{SOM} grid position $i$.

  \item Update learning rate and radius based on respective input decay rates. Similar to learning rate decay, this neighbourhood decay function can be expressed with an exponentially or linearly decaying $\sigma$:

  \begin{equation} 
    \label{eqn:method-som-learing_radius}  
        \alpha(t) = \alpha(0)\sigma
  \end{equation} 

  \item Iterate until a training epoch stop condition is met or learning and neighbourhood rates have decayed to a limit or zero. In our approach, each iteration of the full dataset is considered an epoch as the entire dataset is taken into account with no mini-batch training. 
  
\end{enumerate}

The \ac{SOM} output is a set of learned neuron weight vectors associated to locations on the \ac{SOM} grid. We interpret these vectors and locations using a \acl{UMAT} \citep[UMAT]{ultsch1993self}. This \acs{UMAT} is visualised as a heat map of the Euclidean distance between each neuron and its neighbourhood. We display the learned weights of each neuron by reconstructing the weight vector as an image with the decoder side of the autoencoder. Given an appropriately trained \ac{SOM} will contain weights mapped to the latent vector training set, they can ideally be reconstructed into an approximation of the radio-astronomical features encoded into the neuron weight. We display these weights and the \ac{RGZ} image of the closest matching latent vector on each neuron over the \acs{UMAT}. Additionally, we assess the ability of the \ac{SOM} to separate complexity and anomalies by plotting the distribution of the \acs{UMAT} distance value of each \ac{SOM} neuron and colour coding the closest matching \ac{RGZ} label and source classification of each neuron.

\subsection{K-means Clustering}
We segment the \ac{SOM} \ac{UMAT} in 4 and 8 clusters using the K-means algorithm \citep{lloyd1982least}. This algorithm groups objects by assigning inputs a cluster based on a metric such as Euclidean distance. This is an iterative process where the distance between each cluster pair is calculated as the average distance of its consistent objects. Input clusters are continually refined based on this distance until the changes in each cluster reach a stop condition. These clusters are discrete, where an object is assigned to only one cluster. We use these clusters as proxies for complex and simple feature vectors on the \ac{SOM}. K-means clusters of K = 4 and K = 8 clusterings were chosen solely to demonstrate the general clustering ability of the system with the $20\times20$ sized \ac{SOM} used in this paper. The K-means clustering is conducted on the learned weight vectors of each neuron and was implemented using the Scikit learn package \citep{scikit-learn}.

We display K-means clustering results by colouring each neuron on the \ac{UMAT} grid to indicate its associated cluster. Entropy, $\hat{E}$ is also used here as a metric to describe the distribution labels in the matching data samples (images) of each neuron's receptive field.

    \begin{equation} 
        \label{eqn:entropy}  
        \hat{P_i} =\frac{n_i}{N}
    \end{equation} 
    
    \begin{equation} 
        \label{eqn:entropy}  
        \hat{E} = -\sum_{i}\hat{P_i}log_2\hat{P_i}
    \end{equation} 

where $n_i$ is the number of class occurrences $i$ and $N$ the total number of occurring classes. A low entropy indicates good consensus where most images matching a given neuron have the same label. Conversely, a high entropy indicates the  matching images of a neuron have a wide range of different labels. We normalise this entropy for clarity to a range from 0 to 1.

The complete system outlined in this method section is shown in Figure \ref{fig:ae_architecture_full}.

\section{Results}
This section outlines the results and performance of our approach at each stage of the method.

\subsection{Autoencoder Training and Image Reconstruction}

\begin{figure*}[!ht]
        \includegraphics[trim=1.1cm 0 0 0, scale=0.62]{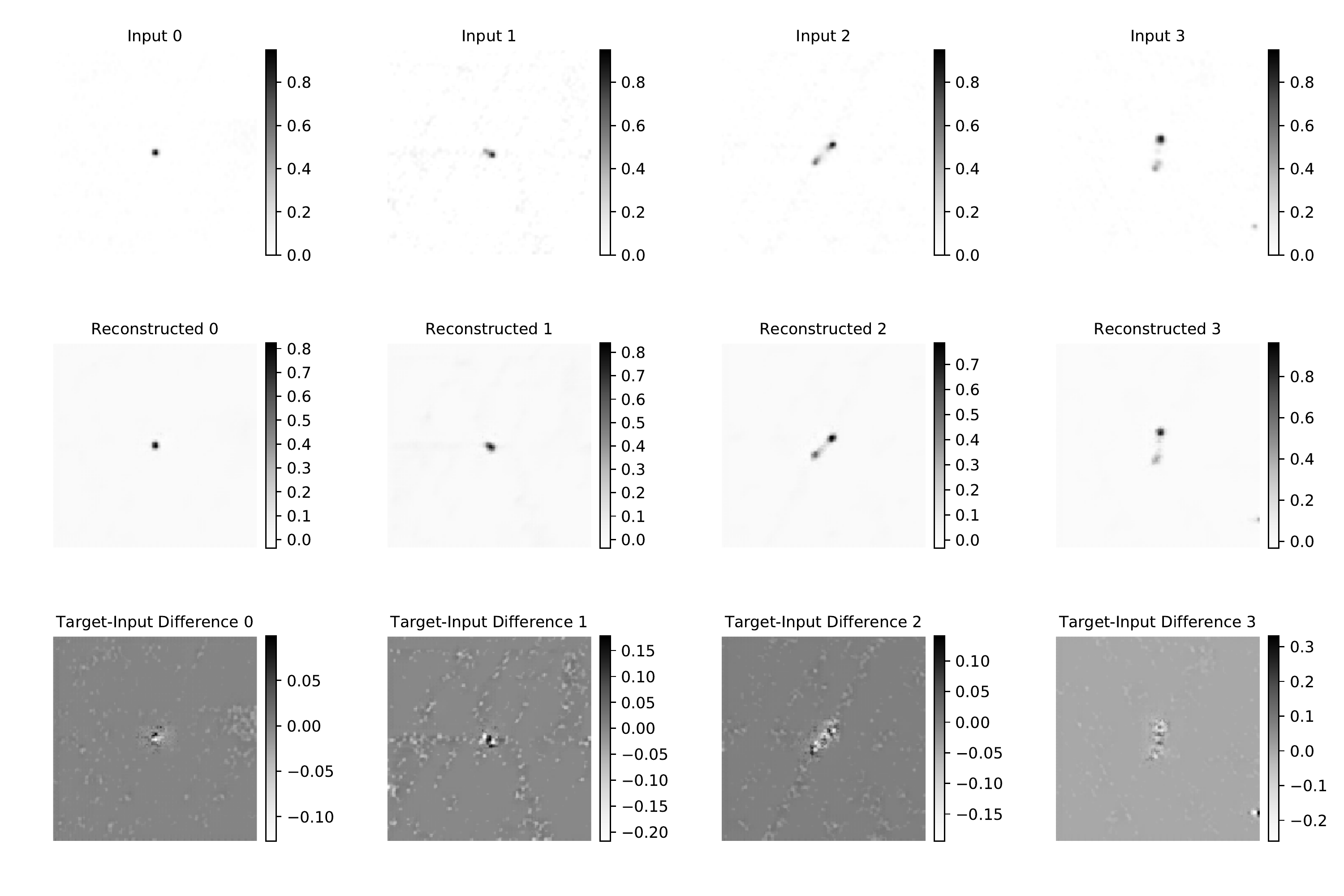}
    \caption{Convolutional autoencoder prediction of \ac{RGZ} input images after 3 training epochs. Top row: Original preprocessed input with pixel intensity scale bars, Middle row: trained autoencoder prediction also with pixel intensity scale bars, Bottom row: Difference image between predicted and original image, with scale bars showing the difference in pixel intensity.  
    \label{fig:ae_results}}
\end{figure*}

  The autoencoder trained on \ac{RGZ} images demonstrates successful compression and decompression across the dataset. This is demonstrated in Figure \ref{fig:ae_results} where the reconstructed image strongly approximates the input image, the general morphology and most of the peak and component counts. From this figure, we determine the autoencoder is capable of recognising and preserving enough key image features to successfully predict the original image from the compressed latent vector. The difference images in the figure show the autoencoder loses most fidelity around the edges of regions and reconstructs background noise with low error. Blurring in the reconstructed image has a square kernel shape and is expected due to the shape of the max pooling and convolutional layers. Autoencoder difference images also show the background noise of each image. Additional layers and training may allow the autoencoder to better generalise the dataset image features to remove this blurring and background noise. 
  
\begin{figure*}
    \centering
        \includegraphics[trim=0 0 0 1.5cm,width=0.55\textwidth]{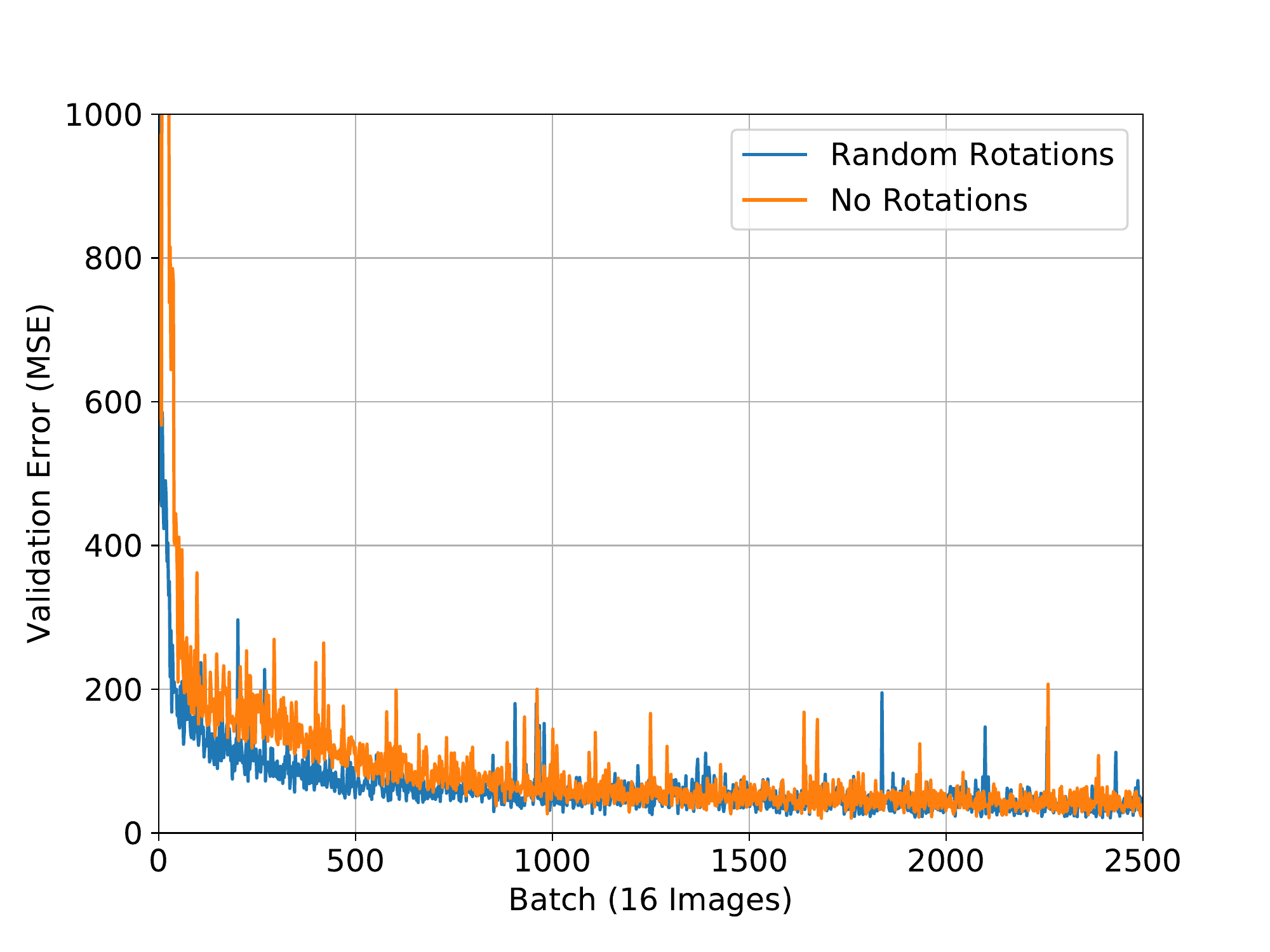}
    \caption{Autoencoder error per batch as mean squared difference between input target image and reconstructed image. \label{fig:error_per_batch}}
\end{figure*}

  Figure \ref{fig:error_per_batch} indicates that training with random rotation augmentations allows the autoencoder to converge on a solution faster and with slightly lower error than training without random rotation augmentations, despite training on the same training set. Faster convergence with the random rotations is apparent early in training even though the network has not yet been trained on more than one rotation per image. This is likely the result of increased variance in the images. This variance may be caused by the rotation of any remnants of the instrument \ac{PSF} which are ordinarily oriented across all images. Similar effects have been observed with increased variance in autoencoder training using random noise injection which has been shown to improve autoencoder training \citep{vincent2010stacked}.
  
  The average training time for this autoencoder using random rotation augmentations and with an un-augmented training set is 26.51 and 25.17 seconds per epoch of 10,000 images respectively. Figure \ref{fig:error_per_batch} suggests that training has converged by the 3\ts{rd} epoch after the 1500\ts{th} batch for both training methods. At this epoch, the total training time is 79.53 seconds and 75.51 seconds for the random rotation and normal training conditions respectively. The slight increase in training time between these two training approaches is negligible and likely the consequence of performing the rotation operation on each training image. The time to encode 30,000 \ac{RGZ} images for \ac{SOM} training or validation is 252.75 seconds at an average of 0.0084 of a second per image. While the random rotation augmentation results in a total encoding time of 440.25 seconds at an average of 0.015 of a second per image.

\subsection{Self Organising Map of Latent Image Vectors}

\begin{figure*}
    \centering
    \includegraphics[width=1.06\textwidth,trim={5cm 6cm 0 6cm},clip]{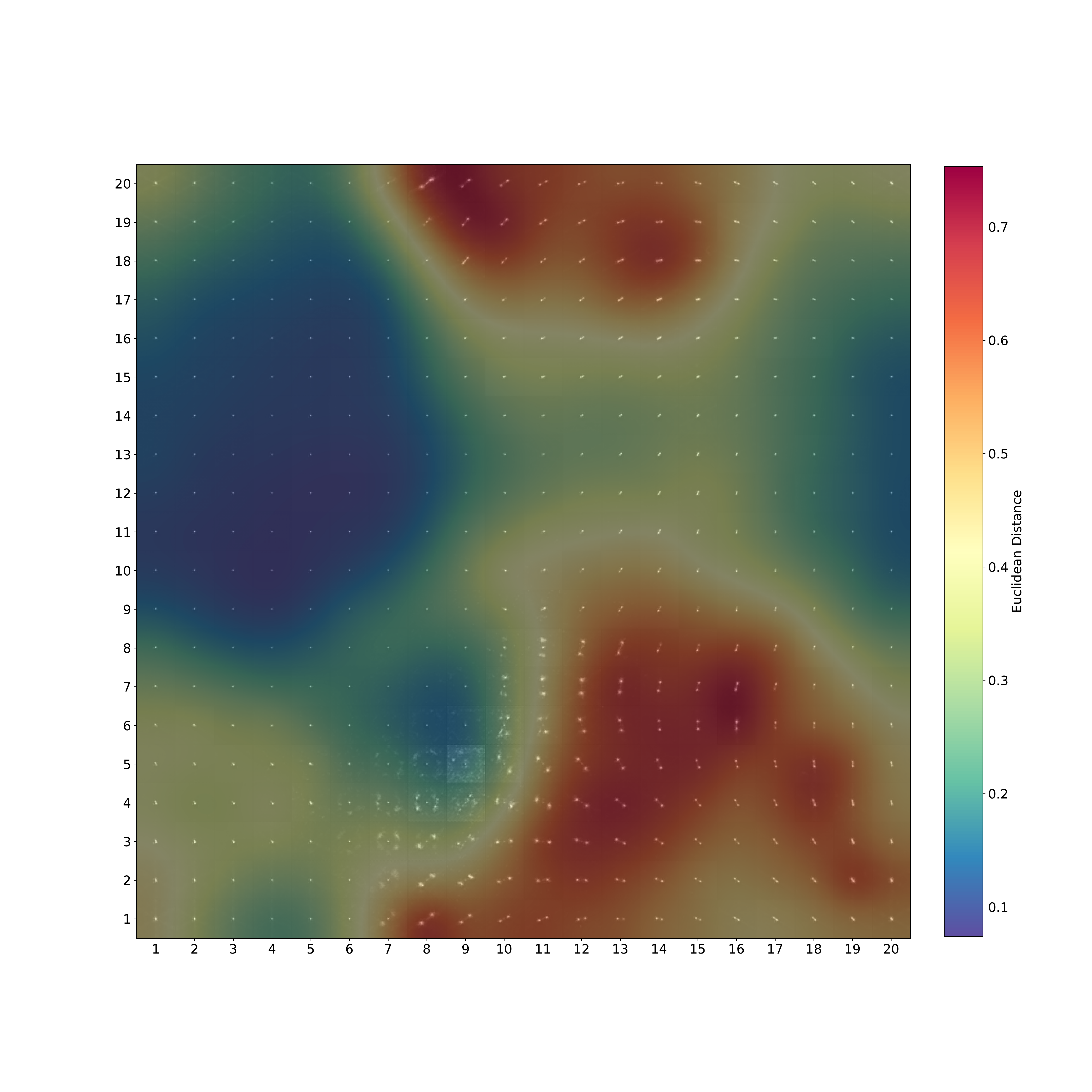}
    \caption{$20\times20$ toroidal \ac{SOM} \ac{UMAT} trained using latent vectors produced by an autoencoder without random rotation training augmentations. Each neuron is displaying false colour decoded neuron weight images at $120\times120$ pixels}. \label{fig:10x10_lgl_decoded_norot}
\end{figure*}

\begin{figure*}
    \centering
    \includegraphics[width=1.06\textwidth,trim={5cm 6cm 0 6cm},clip]{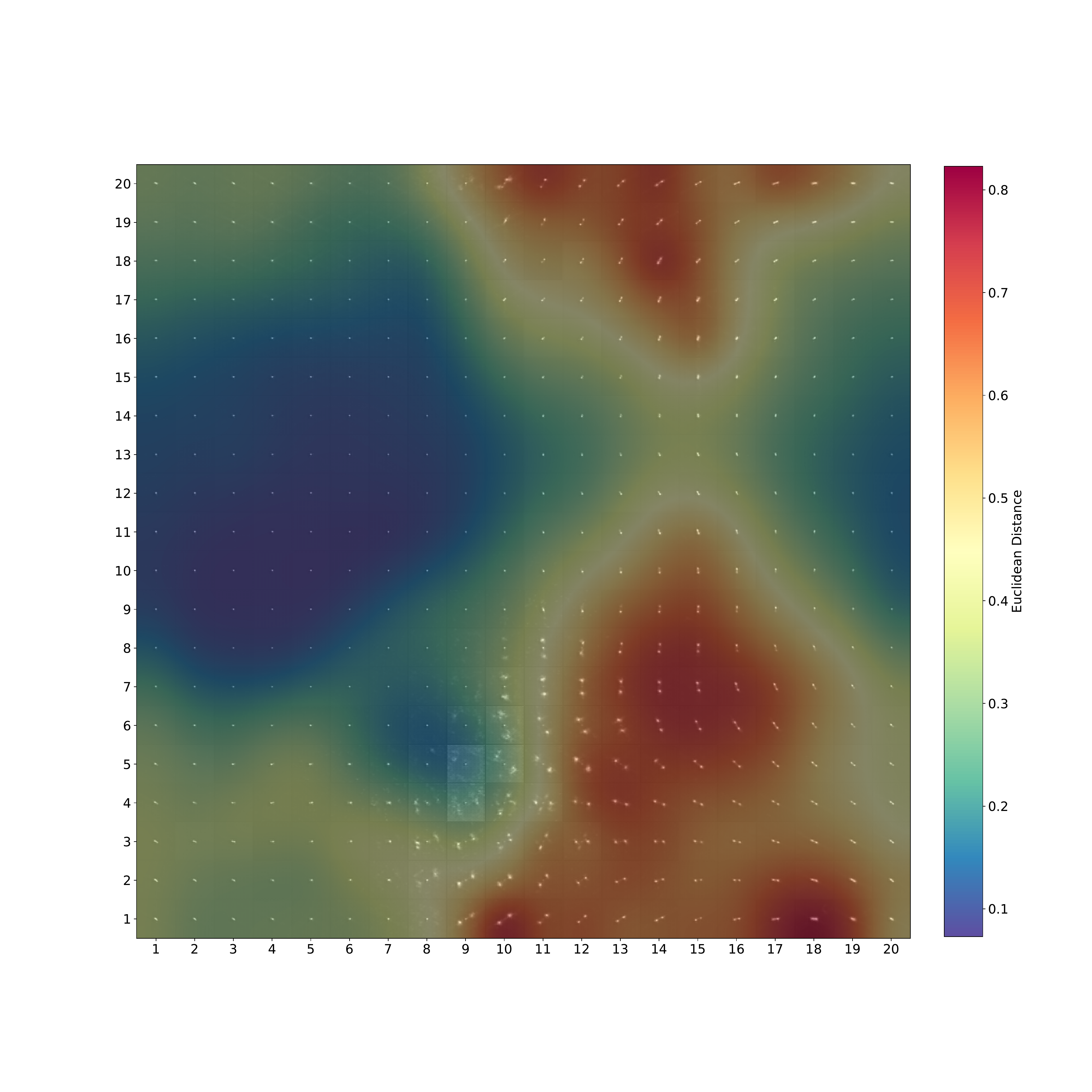}
    \caption{$20\times20$ toroidal \ac{SOM} \ac{UMAT} trained using latent vectors produced by an autoencoder with random rotation training augmentations. Each neuron is displaying false colour decoded neuron weight images at $120\times120$ pixels}. \label{fig:10x10_lgl_decoded_rot}
\end{figure*}

\begin{figure*}
    \centering
    \includegraphics[width=1.06\textwidth,trim={5cm 6cm 0 6cm},clip]{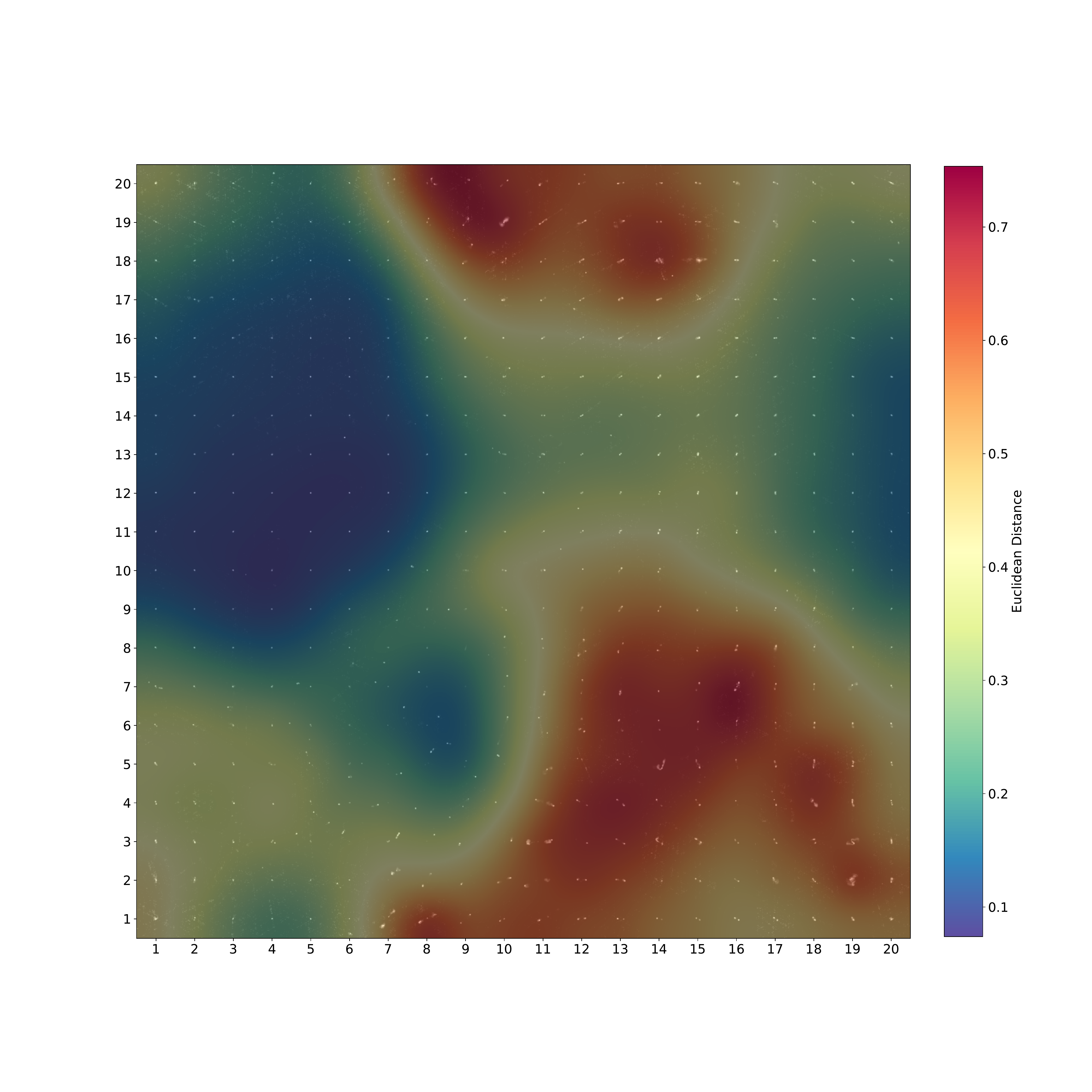}
    \caption{$20\times20$ toroidal \acs{SOM} \ac{UMAT} trained using latent vectors produced by an autoencoder without random rotation training augmentations. Each neuron is displaying the \ac{RGZ} image with the closest matching latent vector transformation to the learned neuron weight. \label{fig:10x10_lgl_rgzmatch_umat_norot}}
\end{figure*}

\begin{figure*}
    \centering
    \includegraphics[width=1.06\textwidth,trim={5cm 6cm 0 6cm},clip]{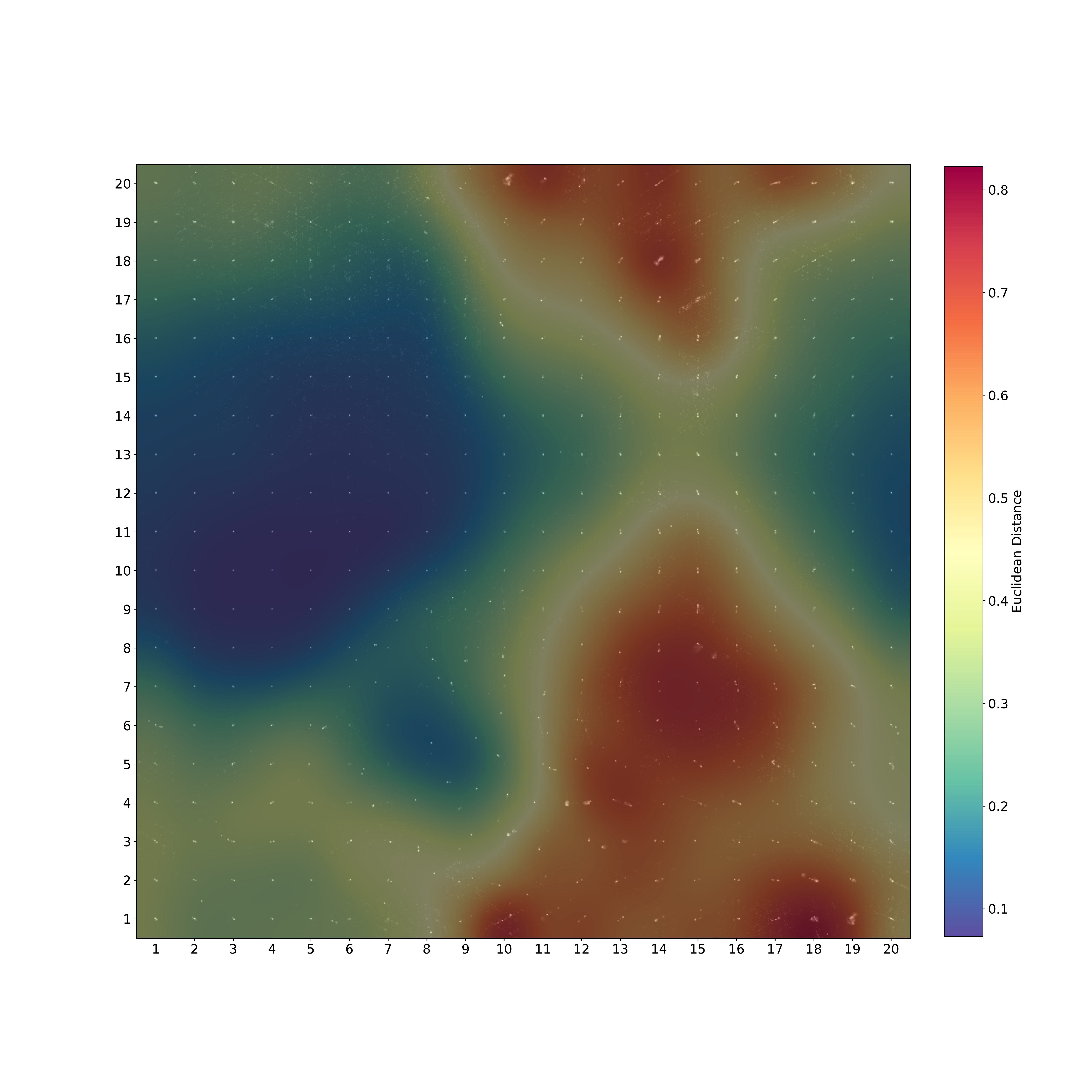}
    \caption{$20\times20$ toroidal \acs{SOM} \ac{UMAT} trained using latent vectors produced by an autoencoder with random rotation training augmentations. Each neuron is displaying the \ac{RGZ} image with the closest matching latent vector transformation to the learned neuron weight. \label{fig:10x10_lgl_rgzmatch_umat_rot}}
\end{figure*}

  The \acl{SOM} \citep{kohonen1997exploration} was trained to produce a $20\times20$ neuron toroidal \ac{UMAT} as displayed in Figures \ref{fig:10x10_lgl_decoded_norot}-\ref{fig:10x10_lgl_rgzmatch_umat_rot}. This map was created in 25.528 seconds with an average of 2.127 seconds per epoch for 12 epochs. We trained the \ac{SOM} using a linear learning rate and neighbourhood decay function, with a Gaussian neighbourhood function. This configuration was chosen as it was the set of training hyper-parameters that provided the most accurate modelling of the \ac{RGZ} images. An initial learning rate of 0.01 was chosen as a default in a manner similar to the selection of the autoencoder learning rate. The initial rate was decayed during training based on the linear decay function toward 0.001. This decay occurred over 12 training epochs, which was found to sufficiently model the training set. These many epochs were found to be sufficient to model the latent vector features and is similar to \citet{geach2012unsupervised}, which also use a nominal 10 epochs for training. 
  
  Both \acp{UMAT} trained on latent vectors with and without random rotation augmentations are shown with an overlay of the decoded neuron weights and each neuron's closest matching \ac{RGZ} images.In both training cases, these clearly show the morphology distribution of \ac{RGZ} image features where the Euclidean distance between the learned weight of each neuron and their neighbouring neurons displayed as a heat map. In these tests, the decoded weight map illustrates which relationships and morphologies have been modelled, while the map containing the closest matching images illustrates the real radio-astronomical features that match the neuron weights.  
 
  The morphological clusters in these maps are not highly discrete with neurons essentially representing a probability distribution of latent feature vectors. These clustered regions are sub-clustered by orientation, with similarly oriented objects clustered together with gradual transitions between classes. We expect to see this gradual transition between classes of images given these objects do not have entirely discrete classifications. In both decoded neuron weight maps, we observe a number of neurons that appear as a rotated average of a central source radio image and extended emission. These morphologies are clear in Figure \ref{fig:10x10_lgl_decoded_norot}, in neurons of the region 5-3,6-9 and Figure \ref{fig:10x10_lgl_decoded_rot}, in a similar region with neurons of the region 7-4,8-10.
  
\begin{figure*}
  \gridline{
            \fig{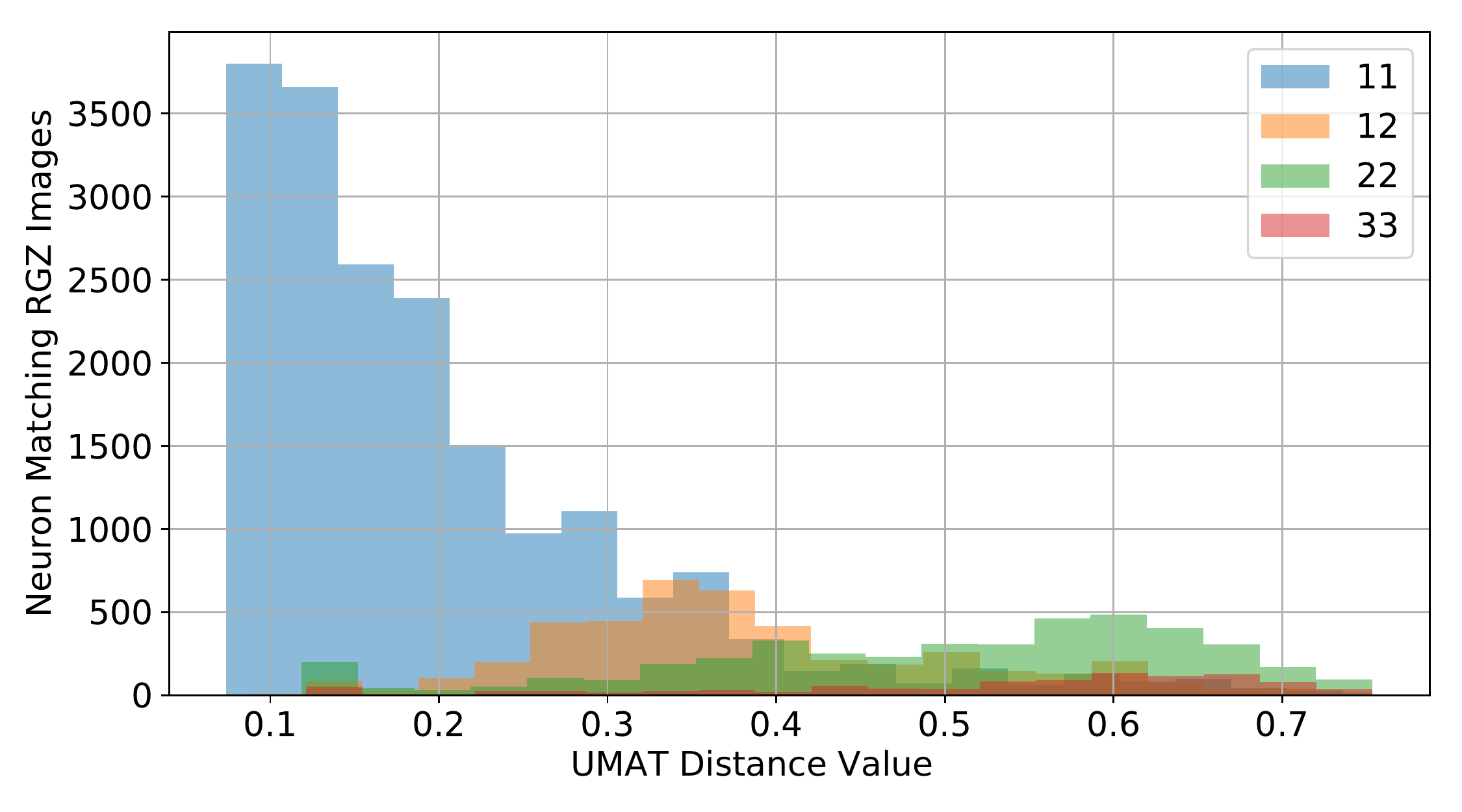}{0.5\textwidth}{(a)}
            \fig{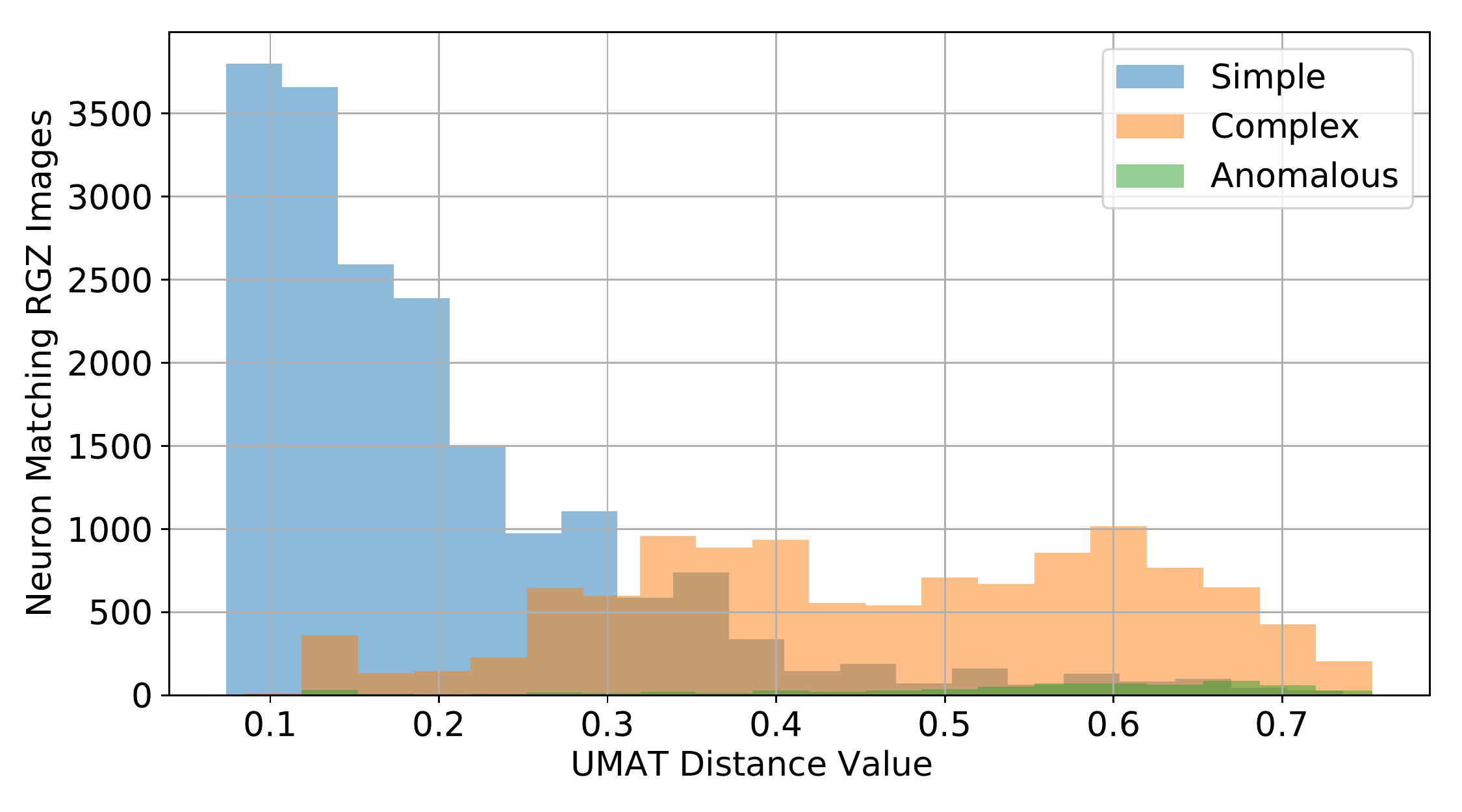}{0.5\textwidth}{(b)}
            }
  \gridline{
            \fig{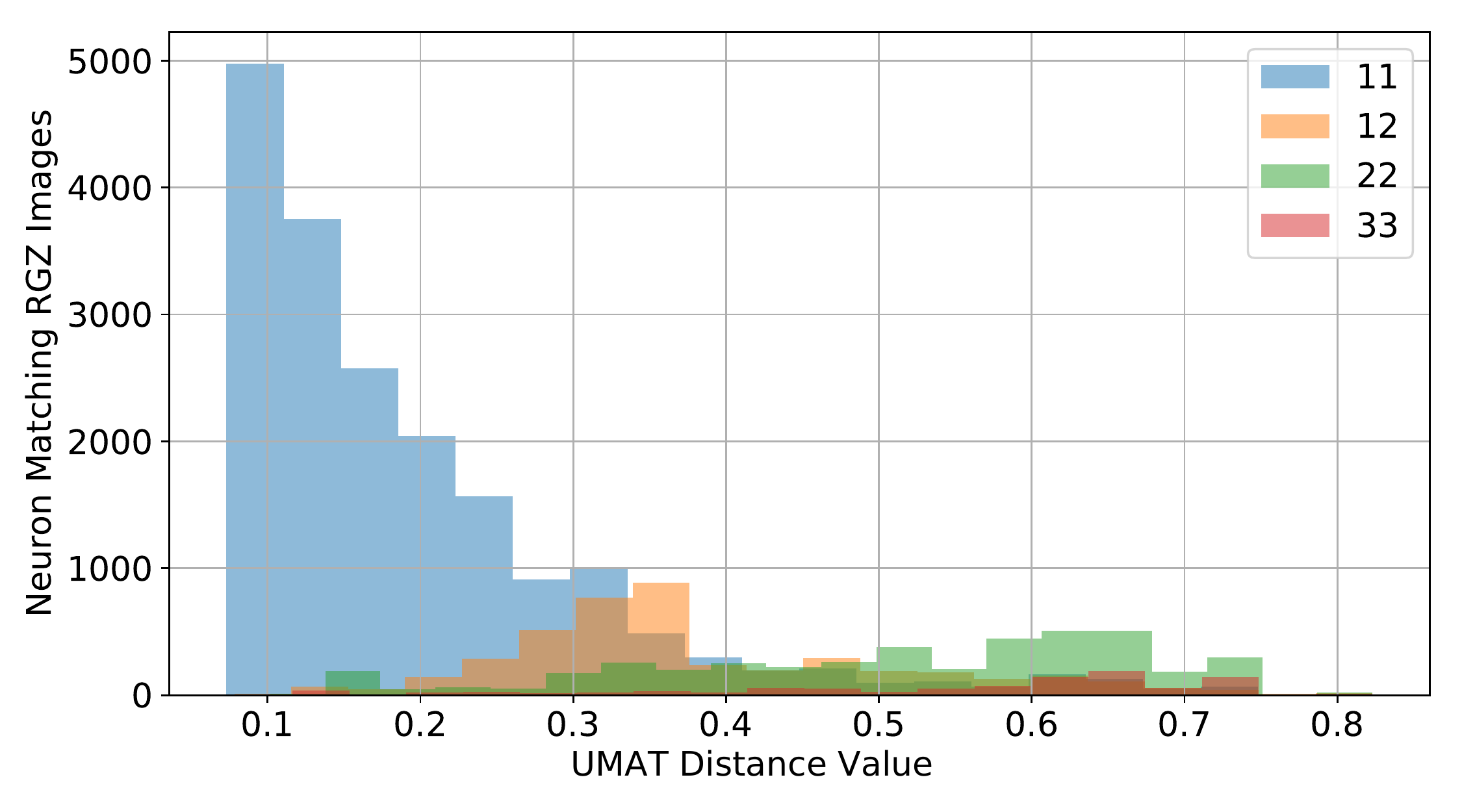}{0.5\textwidth}{(c)}
            \fig{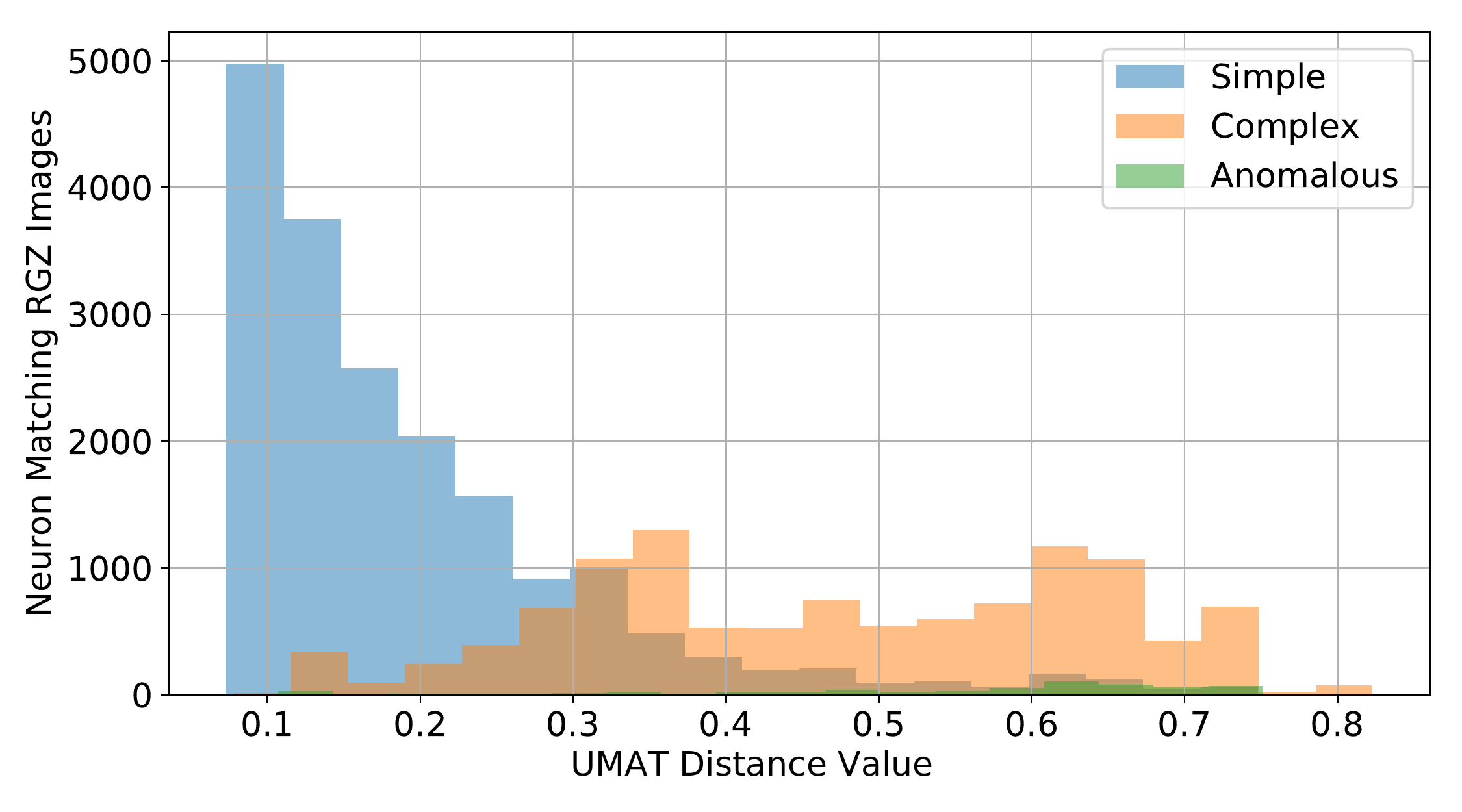}{0.5\textwidth}{(d)}
            }
            
    \caption{Distribution of neuron \ac{UMAT} distance value from the $20\times20$ toroidal \acs{SOM} \ac{UMAT} trained with \ac{RGZ} latent vectors produced by an autoencoder with random rotation training augmentations (a), (b) and without random rotation training augmentations (c), (d). Color coding in (a) and (d) indicates the \ac{RGZ} label of the four most dominant labelled \ac{RGZ} validation images matching each neuron. Coding in (b) and (e) display the class of the labelled source as simple (\ac{RGZ} 11), complex (not \ac{RGZ} 11) and anomalous (\ac{RGZ} label with greater than 3 peaks or components). \label{fig: euc_distance_distributions}}
\end{figure*}

  The low distance regions of the \acp{UMAT} contain prototypes and decoded neuron weights as compact single sources.Morphologies in this central compact region gradually progress in complexity to compact multi-point sources, sources with globular morphologies, and bent-tail sources toward higher distance regions. Images placed in high distance regions have a latent feature vector with a high \ac{UMAT} distance value to surrounding neighbours, which highlight outliers within the \ac{RGZ} image set. These results are also illustrated in Figure \ref{fig: euc_distance_distributions}, which show the distribution of \ac{UMAT} distance values of neurons across the map. This figure shows a clear separation of simple, complex and anomalous classed sources, in addition to \ac{RGZ} label 11, 12, 22 and 33, based on the \ac{UMAT} distance value.

  The differences between the results obtained when trained on the rotated augmentation latent vectors and with normal latent vectors are minor, with the random rotations producing slightly more defined peaks in the \ac{UMAT} distance value distributions in Figure \ref{fig: euc_distance_distributions}. However, it appears that the \ac{SOM} trained on latent vectors produced by an autoencoder with random rotation contains less prominent average rotation weights. This improvement is likely due to the autoencoder encoding rotation information in the latent vector, which has allowed the \ac{SOM} to separate these rotations and clean these average rotated weights. Although rotation is not a meaningful radio-astronomical feature, these changes produce an easier to interpret map for both the decoded weight map and the map displaying the closest matching \ac{RGZ} neuron images. The map trained using latent vectors produced from random rotations in the autoencoder was used for all subsequent tests due to these improvements.
  
    The increased rotational dependency observed in these tests raise questions regarding the true nature of rotational invariance in a \ac{SOM}. For a neuron to be rotationally invariant, morphological features must be the only feature that is clustered by the system. For this to be the case, genuine rotational invariance would result in all neurons on the \ac{SOM} being mapped with the same position angle, or for each neuron to contain all possible position angles, as the observed rotated average morphology seen in decoded neuron images. These concepts suggest that greater rotational invariance may be found on maps with more of the observed averaged rotation neurons.

\subsection{K-means Clustering and Verification with \ac{RGZ} labels}

Test results for the map segmentation using K-means clustering are shown for each \ac{SOM} using an image of the \ac{SOM} grid with decoded neuron weights and map displaying the closest matching \ac{RGZ} image, with a K-means colour coding on each neuron for the assigned K-means cluster \ac{ID} number. All K-means cluster \ac{ID} numbers are arbitrary as they are assigned in an unsupervised manner. All associated ID colours are assigned as discrete colour intervals to visually differentiate individual clusters \acp{ID}. Two tables are included for each test. The Table \ref{table: 20x20_k4_lgl_overall_hc_population_rot} and \ref{table: 20x20_k8_lgl_overall_hc_population_rot_detailed} describe the division of the map assigned to each cluster and associated entropy statistics. Tables \ref{table: 20x20_k4_lgl_overall_hc_population_rot_detailed} and \ref{table: 20x20_k8_lgl_overall_hc_population_rot_detailed} describe for every cluster, the division of neurons with the label of the closest matching \ac{RGZ} image to each neuron, the total division of the \ac{RGZ} images with that label in the cluster and the associated entropy statistics.

These results demonstrate that the K-means clustering is separating \ac{SOM} neurons closely related to morphology and to a lesser extent, the rotation angle of the source features. Clusters also appear to segment morphologies by their relative complexity. There are definitive simple, complex and intermediate groups divided by the clustering, with clear groups of relatively simple clusters which are comprised of mostly point sources (\ac{RGZ} label 11), compact multi-peak single component sources (\ac{RGZ} label 12), complex sources with highly separated sources or sparse sources with distant companions. Regarding general clustering quality, all tests show reasonable connectedness with few neuron clusters inter-mixing. The total clustering time for this \ac{SOM} is negligible at 0.160 and 0.176 seconds for K = 4 and K = 8 clusterings respectively

\begin{figure*}
    \centering
    \includegraphics[width=1.06\textwidth,trim={5cm 6cm 0 6cm},clip]{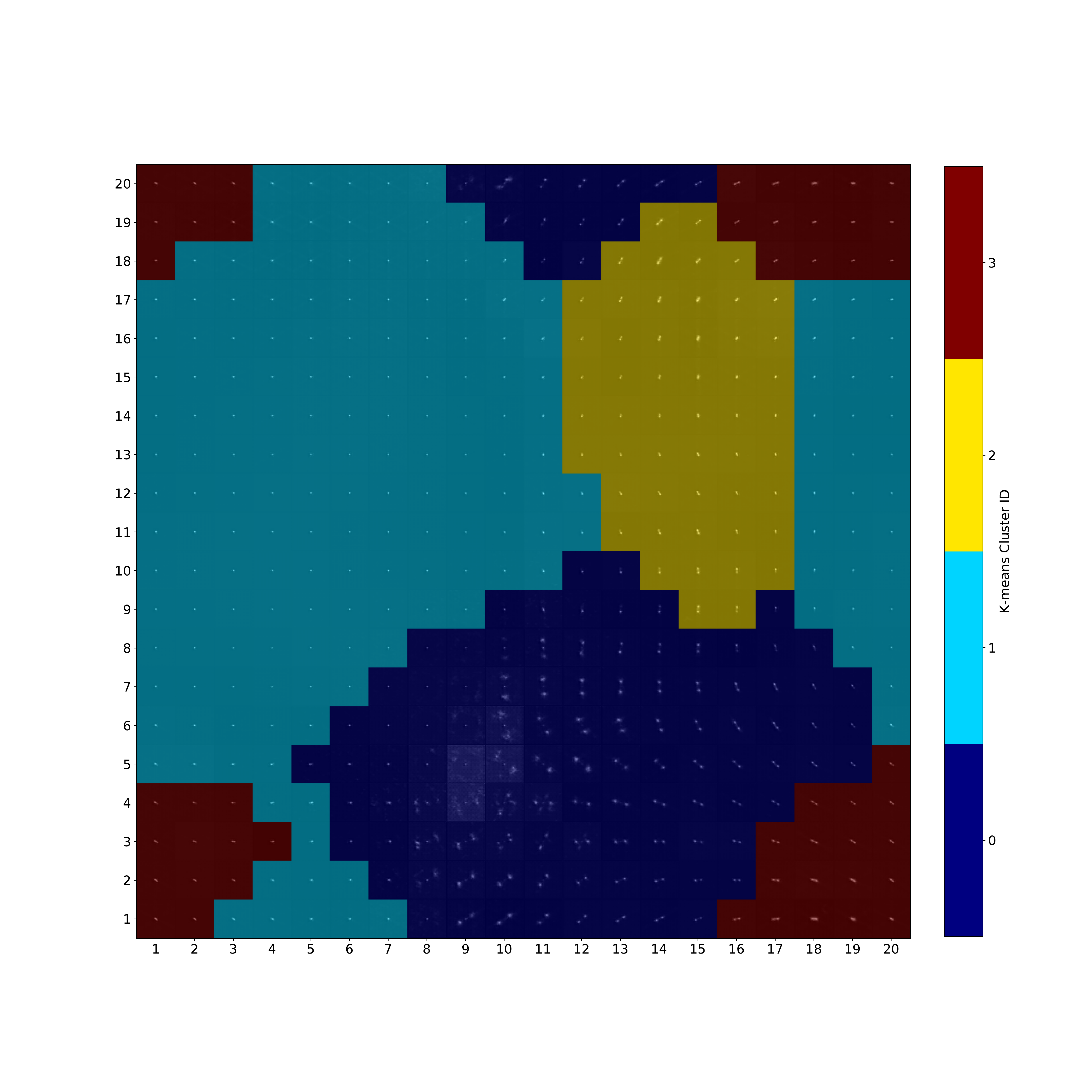}
\caption{$20\times20$ toroidal \acs{SOM}, displaying an overlay of decoded neuron weights with 4 colour coded K-means clusters.}\label{fig:20x20_k4_lgl_decoded_hc_rot}
    \label{fig:my_label}
\end{figure*}

\begin{figure*}
    \centering
    \includegraphics[width=1.06\textwidth,trim={5cm 6cm 0cm 6cm},clip]{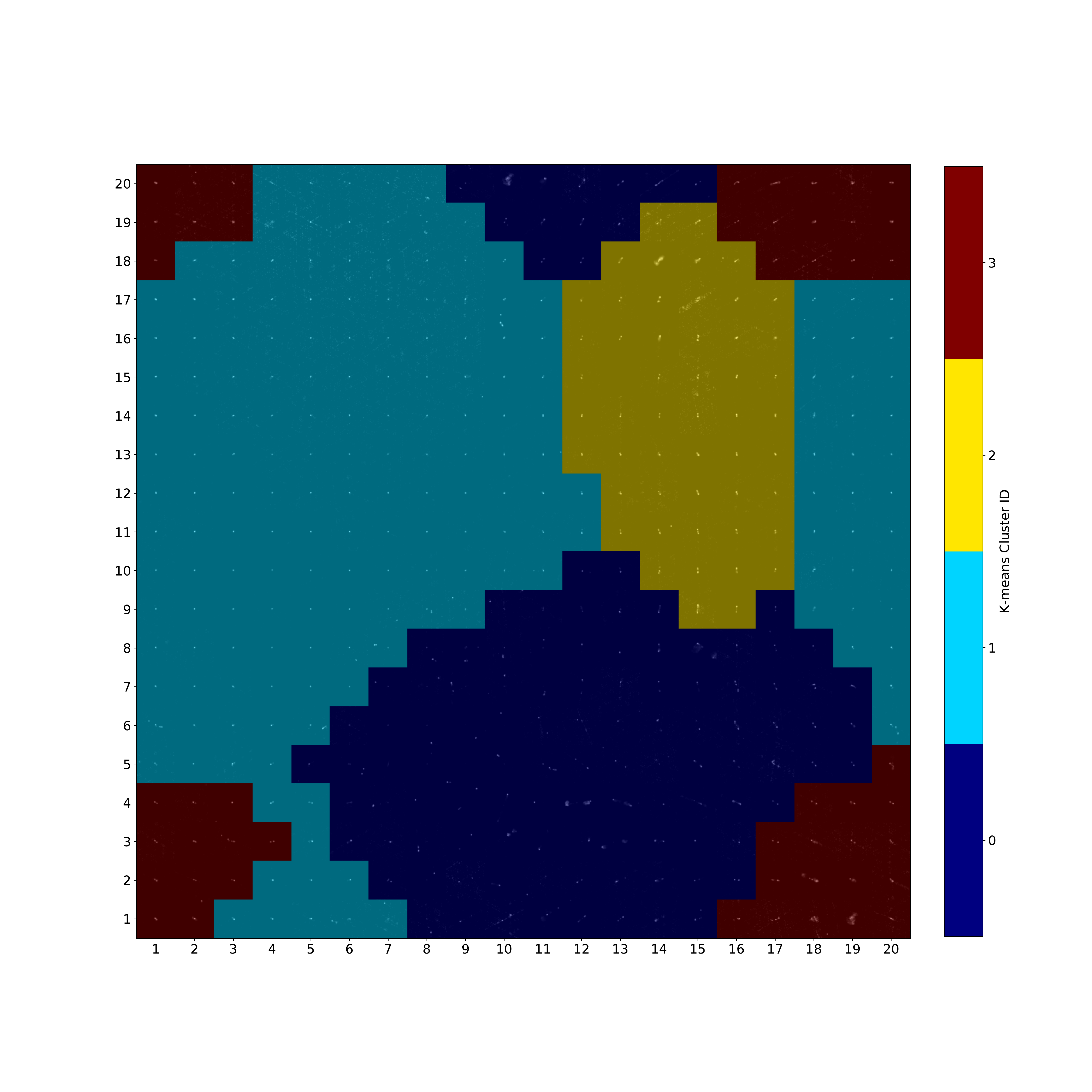}
    \caption{$20\times20$ toroidal \acs{SOM}, displaying the closest matching \acs{RGZ} images with 4 colour coded K-means clusters.\label{fig:20x20_k4_lgl_rgzmatch_umat_hc_rot}}
\end{figure*}

In the K = 4 clusters, the \ac{UMAT} in Figures \ref{fig:20x20_k4_lgl_decoded_hc_rot} and \ref{fig:20x20_k4_lgl_rgzmatch_umat_hc_rot}, appears to be segmented into four groups with varying complexity. As summarised in Table \ref{table: 20x20_k8_lgl_overall_hc_population_rot_detailed}, in the most simple group, cluster \ac{ID} 1, a 0.88 division of the cluster contains \ac{RGZ} labelled point sources, which contains a division of 0.73 of all \ac{RGZ} point sources in the dataset. Similar clustering is seen in the more complex cluster 0, where a vast majority of the cluster contains radio-doubles and a majority of the radio-doubles in the dataset reside. The remaining clusters 2 and 3, appear to segment largely medium complexity sources such as \ac{RGZ} labelled 12 and a mix of 11 and 22 labels.

\begin{figure*}
    \centering
    \includegraphics[width=1.06\textwidth,trim={5cm 6cm 0 6cm},clip]{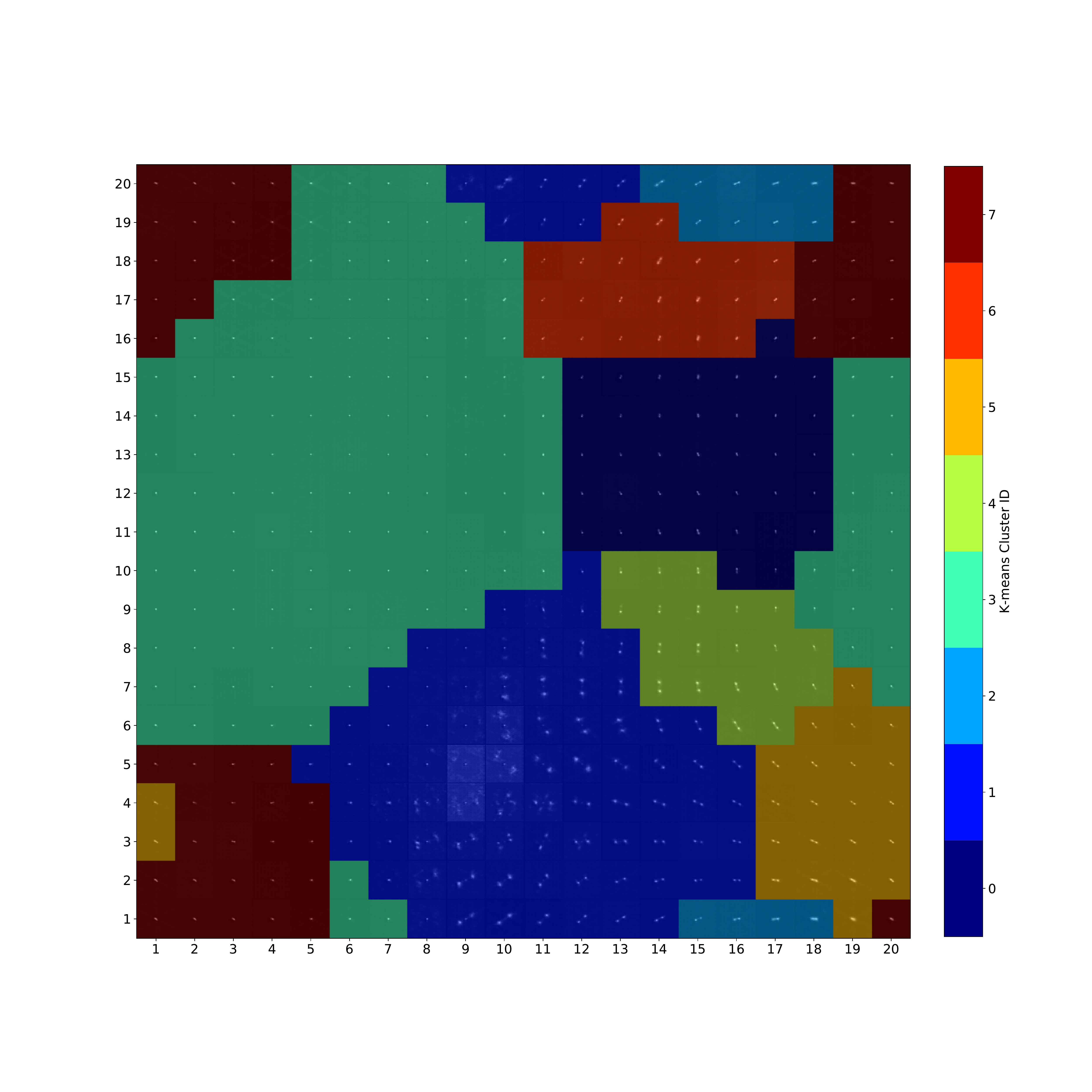}
\caption{$20\times20$ toroidal \acs{SOM}, displaying an overlay of decoded neuron weights with 8 colour coded K-means clusters.}\label{fig:20x20_k8_lgl_decoded_hc_rot}
    \label{fig:my_label}
\end{figure*}

\begin{figure*}
    \centering
    \includegraphics[width=1.06\textwidth,trim={5cm 6cm 0 6cm},clip]{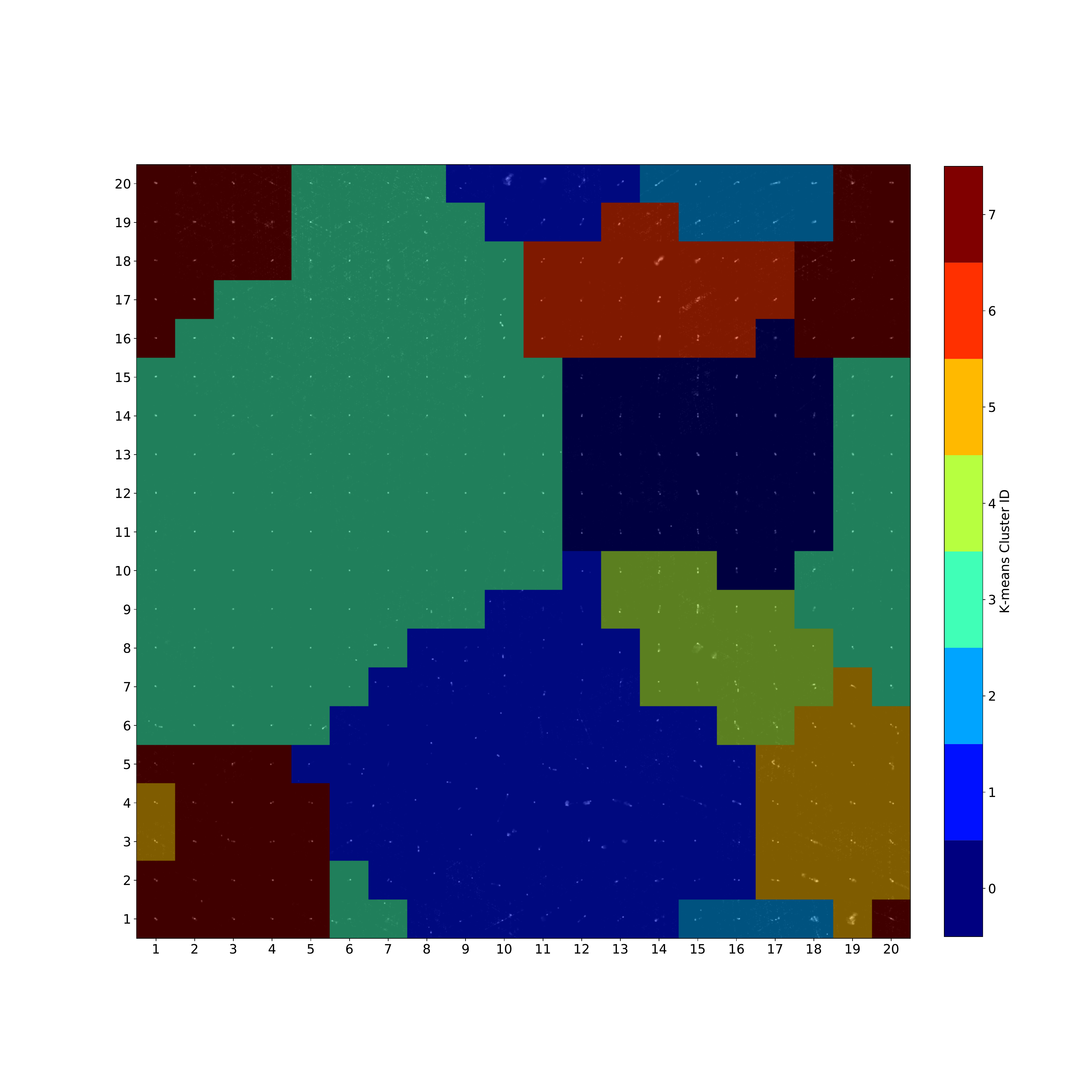}
    \caption{$20\times20$ toroidal \acs{SOM}, displaying the closest matching \acs{RGZ} images with 8 colour coded K-means clusters.\label{fig:20x20_k8_lgl_rgzmatch_umat_hc_rot}}
    \label{fig:my_label}
\end{figure*}

We observe more meaningful clusters with the K = 8 clustering tests shown in Figure \ref{fig:20x20_k8_lgl_decoded_hc_rot} and \ref{fig:20x20_k8_lgl_rgzmatch_umat_hc_rot}. These clusters segment the map into similar groups to the K = 4 clusters but with clusters containing higher population divisions, where many groups are dominated by divisions in some cases of between 0.600 and 0.977 of the same \ac{RGZ} labelled source. This can be seen with complex clusters such as 2 and 4 containing mostly \ac{RGZ} labelled 22 sources. Similarly, simple clusters such as cluster 3, are almost entirely comprised of \ac{RGZ} label 11 point sources and contain most of the point sources from the dataset. Similar to the K = 4 cluster tests, there are a number of intermediate clusters such as cluster 5 and 6, with medium complexity that contain a vast majority \ac{RGZ} label 12 sources.

In the K = 4 and K = 8 clustering tests, neither tables are listing more complex labels such as 33, as these form a minute population of the dataset, but are visible by their morphology in the learned \ac{UMAT}, as contained in the identified highly complex clusters. Most notable are clusters 0 and 3 from the K = 4 cluster test, and clusters 1 and 2 in the K = 8 cluster tests, which contain highly complex and interesting sources showing a wide range of extended features. Although the K = 8 cluster test contains more meaningful clusters, it appears a balance must be reached with the number of clusters and neurons available in the map. Using too many clusters may cause groupings to split logical classes to satisfy the K cluster value, or too few clusters which may result in true divisions and relationships on the map being under-represented or not revealed.

It is evident through these observations and the population division tables, that the K-means algorithm is segmenting morphologies and associating relationships not entirely correlated to the \ac{RGZ} labels despite the clustered maps showing reasonably clear and logically assigned clusters based on morphology and complexity. These results indicate that the relationships learned by the system may be more complex than peak and component counts and therefore under-represent them. This detection of labels different from those of the training set is wholly expected since the system is trained in an unsupervised manner.

Across all tests, there are several evident entropy and map population effects. Most notably, mean cluster entropy values increase with the presence of more complex and anomalous sources. As expected, point sources have the lowest mean entropy due to their simplicity and the greatest neuron population across the map due to the point source bias in the \ac{RGZ} dataset. Higher complexity sources, represent a smaller part of the training set and appear to have consistently smaller cluster populations, but significantly greater mean entropy due to their complexity. These population and entropy statistics effectively reveal not only the types and complexity of morphologies in the dataset but also effectively describe the original label distributions and biases.

The difficulty when trying to use clusters as classifications on a continuous manifold can be seen by both the label cross-overs found between many K-means classes and the point made by \citet{kohonen1997exploration}; \acp{SOM} are not explicitly designed for hard classification. The original principles of \ac{SOM} learning will not produce highly distinct clusters, but will instead produce these results: a semantic map of outliers, regions, and morphologies rather than highly distinct groups. These qualities are largely seen with the blurring of features in neuron weights due to the relatively small $20\times20$ map size compressing the true feature space. It is possible that in exceptionally large \ac{SOM} these relationships may have enough space to become sufficiently separated for discrete classification. Reduction of blurring effects in small crowded \acp{SOM} such as ours and the highlighting of outliers have also been successfully shown in \citet{tasdemir2009exploiting}, by instead implementing a new connectivity measure for the similarity of \ac{SOM} prototypes that produces a more effective detection of manifold structures.

The total training time of this system is competitive with other methods from literature such as \citet{polsterer_parallelized_2016}. Our approach produces similar \ac{SOM} morphologies with square neurons and significantly reduced processing time as python code using a 24 core \ac{CPU} requiring 14.55 minutes for the full autoencoder training of 10,000 images, encoding of 10,000 images, \ac{SOM} training of 30,000 images, validation of an additional 30,000 images and final clustering, compared to 17 days with 200,000 images using python code on a 8 core \ac{CPU}. The difference in data volume is the likely cause, where even with random rotation autoencoder training, our \ac{SOM} training latent vectors contain only 900 elements per image, opposed to a total of 1,331,280 elements per image including rotations used in \citet{polsterer_parallelized_2016}.

 \begin{table*}[h!]
  \begin{center}
    \caption{Cluster population and entropy statistics for the $20\times20$ toroidal \acs{SOM} \acs{UMAT} with K = 4 clusters.}
        \label{table: 20x20_k4_lgl_overall_hc_population_rot}
    \begin{tabular}{lcccc}
    \hline
      K-means  &   Cluster Population  &  Minimum  &  Maximum &  Mean \\
      Cluster&               Over Map              &    Entropy          &     Entropy         &  Entropy             \\
      \hline

0              &                       0.278 &         0.00 &         1.00 &          0.46 \\
1              &                        0.460 &         0.00 &         0.70 &          0.23 \\
2              &                       0.118 &         0.04 &         0.98 &          0.52 \\
3              &                     0.145 &         0.05 &         0.84 &          0.43 \\

      \hline
     \end{tabular}
  \end{center}
\end{table*}


\begin{table*}[h!]
  \begin{center}
    \caption{Divisions of clusters based on the label of the closest matching \ac{RGZ} image, matching clustered label divisions and entropy statistics for each of the K = 4 clusters in the $20\times20$ toroidal \acs{SOM} \acs{UMAT}}
        \label{table: 20x20_k4_lgl_overall_hc_population_rot_detailed}
    \begin{tabular}{llccccc}
    \hline
         Cluster & \ac{RGZ} & Division of & Division of   & Mean & Minimum & Maximum \\
        ID & Label & Matching Label & All Image    & Entropy & Entropy & Entropy \\
        & & In Cluster & Labels In & & &\\
        & & & Cluster &&& \\
       \hline 

0 & 22         &                          0.622 &                                0.742 &          0.61 &         0.12 &         1.00 \\
  & 11         &                          0.369 &                                0.186 &          0.21 &         0.00 &         0.52 \\
    \hline
1 & 11         &                           0.88 &                                0.733 &          0.21 &          0.0 &          0.3 \\
  & 12         &                           0.12 &                                0.256 &          0.38 &          0.2 &          0.7 \\
    \hline
2 & 12         &                          0.511 &                                0.279 &          0.50 &         0.15 &         0.75 \\
  & 22         &                          0.404 &                                0.204 &          0.63 &         0.34 &         0.98 \\
  & 11         &                          0.085 &                                0.018 &          0.21 &         0.04 &         0.48 \\
    \hline
3 & 12         &                          0.672 &                                0.453 &          0.52 &         0.18 &         0.84 \\
  & 11         &                          0.241 &                                0.063 &          0.12 &         0.05 &         0.39 \\
  & 22         &                          0.086 &                                0.054 &          0.64 &         0.51 &         0.78 \\
  
        \hline
     \end{tabular}
  \end{center}
\end{table*}

 \begin{table*}[h!]
  \begin{center}
    \caption{Cluster population and entropy statistics for the $20\times20$ toroidal \acs{SOM} \acs{UMAT} with K = 8 clusters.}
        \label{table: 20x20_k8_lgl_overall_hc_population_rot}
    \begin{tabular}{lcccc}
    \hline
      K-means  &   Cluster Population  &  Minimum  &  Maximum &  Mean \\
      Cluster&               Over Map              &    Entropy          &     Entropy         &  Entropy             \\
      \hline

0 &                                   0.125 &         0.03 &         0.84 &          0.26 \\
1 &                                0.210 &         0.00 &         1.00 &          0.42 \\
2 &                                    0.042 &         0.15 &         0.74 &          0.62 \\
3 &                                   0.330 &         0.00 &         0.31 &          0.23 \\
4 &                                0.055 &         0.09 &         0.82 &          0.48 \\
5 &                                  0.055 &         0.39 &         0.69 &          0.58 \\
6 &                                 0.050 &         0.36 &         0.98 &          0.62 \\
7 &                                   0.132 &         0.02 &         0.78 &          0.35 \\

      \hline
     \end{tabular}
  \end{center}
\end{table*}


\begin{table*}[h!]
  \begin{center}
    \caption{Divisions of clusters based on the label of the closest matching \ac{RGZ} image, matching clustered label divisions and entropy statistics for each of the K = 8 clusters in the $20\times20$ toroidal \acs{SOM} \acs{UMAT}}
        \label{table: 20x20_k8_lgl_overall_hc_population_rot_detailed}
    \begin{tabular}{llccccc}
    \hline
         Cluster & \ac{RGZ} & Division of & Division of   & Mean & Minimum & Maximum \\
        ID & Label & Matching Label & All Image    & Entropy & Entropy & Entropy \\
        & & In Cluster & Labels In & & &\\
        & & & Cluster &&& \\
       \hline

0 & 12 &       0.545 &                                0.140 &          0.59 &         0.39 &         0.69 \\
  & 22 &       0.409 &                                0.097 &          0.58 &         0.48 &         0.67 \\
  & 11 &       0.045 &                                0.005 &          0.39 &         0.39 &         0.39 \\
      \hline
1 & 22 &       0.548 &                                0.495 &          0.61 &         0.12 &         1.00 \\
  & 11 &       0.452 &                                0.172 &          0.20 &         0.00 &         0.51 \\
    \hline
2 & 22 &       0.900 &                                0.194 &          0.62 &         0.36 &         0.98 \\
  & 11 &       0.050 &                                0.005 &          0.48 &         0.48 &         0.48 \\
  & 12 &       0.050 &                                0.012 &          0.75 &         0.75 &         0.75 \\
\hline
3 & 11 &       0.977 &                                0.584 &          0.22 &         0.00 &         0.30 \\
  & 12 &       0.023 &                                0.035 &          0.29 &         0.28 &         0.31 \\
    \hline
4 & 22 &       0.824 &                                0.151 &          0.67 &         0.55 &         0.74 \\
  & 11 &       0.176 &                                0.014 &          0.37 &         0.15 &         0.52 \\
  \hline
5 & 12 &       0.660 &                                0.407 &          0.45 &         0.15 &         0.75 \\
  & 11 &       0.302 &                                0.072 &          0.11 &         0.02 &         0.17 \\
  & 22 &       0.038 &                                0.022 &          0.56 &         0.34 &         0.78 \\
\hline
6& 12 &       0.682 &                                0.174 &          0.53 &         0.18 &         0.82 \\
  & 22 &       0.182 &                                0.043 &          0.61 &         0.49 &         0.78 \\
  & 11 &       0.136 &                                0.014 &          0.11 &         0.09 &         0.12 \\
    \hline
7& 11 &       0.600 &                                0.136 &          0.14 &         0.03 &         0.26 \\
  & 12 &       0.400 &                                0.233 &          0.43 &         0.25 &         0.84 \\
    \hline  

     \end{tabular}
  \end{center}
\end{table*}

\clearpage

\section{Future Work}
Our plans are to improve each of the three distinct components, the autoencoder, self-organising map and clustering algorithms. By varying the latent vector sizes and structure of the autoencoder we will achieve a better balance between training time and accuracy. Additionally, we will be using a stacked architecture to train latent vectors for multi-channel data. The \ac{SOM}  will be further improved with heat map display of entropy in addition to gathering more performance metrics such as precision and reliability. We will also investigate other clustering algorithms trained in different learning spaces and projections. Additional variables such as map size can be eliminated and more dynamic relationships examined using a growing \ac{SOM} \citep{rauber2002growing}. We aim to further investigate affine invariance in \ac{SOM} by training on images aligned to a common major axis and with all central components scaled to the same size. As previously discussed, the \ac{SOM} output appears is continuous. Consequently, the challenge of more definitive and in-depth source classification can be viewed as a regression problem. We aim to continue working in this direction to create a machine learning regression framework auxiliary to the \ac{SOM} to regress the continuous \ac{SOM} morphologies into discrete classes. Additionally, future investigations will also focus on determining the scalability of this method when applied to significantly increased data volume and more numerous classes.

\section{Conclusion}

We conclude that the coupling of self-organising maps with convolutional autoencoders is an effective method of data exploration and unsupervised clustering of radio-astronomical images. Our approach directly addresses the growing survey processing time and provides a better means to explore large datasets automatically with a total processing time less than 15 minutes for 80,000 images. Our results demonstrate an accurate visualisation of morphology distributions found within the \ac{RGZ} dataset. Our results show the capabilities of this method in locating outliers as high \ac{UMAT} distance values and in K-means clustering with a distinct class of highly complex sources with low dataset population. By combining clustering with citizen science projects such as Radio Galaxy Zoo, greater efficiency can be achieved with volunteers inspecting only a small sample of objects from each cluster or being guided by likely morphologies in each cluster. The speed of this method holds implications for use on large future surveys, large-scale instruments such as the \ac{SKA} \citep{johnston_science_2007} and in other big data applications.  \\

\section*{Acknowledgements}
The authors acknowledge the Radio Galaxy Zoo Project builders and volunteers listed in full at \\http://rgzauthors.galaxyzoo.org for their contribution to \ac{RGZ} dataset and labels used in this paper. We also acknowledge the National Radio Astronomy Observatory (NRAO) and the Karl G. Jansky Very Large Array (VLA) as the source of this radio data. Partial support for L.R is provided by the U.S National Science Foundation grant AST17-14205 to the University of Minnesota. H.A benefited from grant DAIP \#066/2018 of Universidad de Guanajuato. 

\bibliographystyle{aasjournal}
\bibliography{aesom}

\end{document}